\documentclass[sigconf]{acmart}

\AtBeginDocument{%
  }

\setcopyright{none} 
\copyrightyear{2025}
\acmYear{2025}
\acmDOI{10.1145/3695053.3731056}

\acmConference['ISCA 2025']{The 52nd IEEE/ACM International Symposium on Computer Architecture}{June 21--25, 2025}{Tokyo, Japan}

\acmISBN{979-8-4007-1261-6/2025/06}

\renewcommand\footnotetextcopyrightpermission[1]{} 

\usepackage{makecell}

\usepackage[normalem]{ulem}

\usepackage[most]{tcolorbox} 

\usepackage{tikz}
\usepackage{eso-pic}
\usepackage{fancyhdr}

\begin{document}

\title{NMP-PaK: Near-Memory Processing Acceleration of Scalable \textit{De Novo} Genome Assembly}


\author{\normalsize{Heewoo Kim\textdagger, Sanjay Sri Vallabh Singapuram*, Haojie Ye*, Joseph Izraelevitz\textdagger, Trevor Mudge*, Ronald Dreslinski*, Nishil Talati*
\\ {\textdagger}University of Colorado, Boulder, CO, USA; *University of Michigan, Ann Arbor, MI, USA\\ Email: heewoo.kim@colorado.edu}}

\renewcommand{\shortauthors}{}



\newcommand{\NMPGAIN}{16$\times$}
\newcommand{\CPUNMPGAIN}{2.6$\times$}
\newcommand{\NMPONLYGAIN}{6.2$\times$}

\newcommand{\NMP}{NMP-PaK}
\newcommand{\CPUNMP}{CPU-PaK}
\newcommand{\CPU}{CPU baseline}

\newcommand{\NMPPE}{\NMP\ with ideal PE}
\newcommand{\NMPFW}{\NMP\ with ideal forwarding logic}

\newcommand{\NMPFWGAIN}{18.2$\times$}
\newcommand{\FWGAIN}{1.14$\times$}
\newcommand{\FWTRAFFIC}{18\%}

\newcommand{\THROUGHPUT}{8.3$\times$}

\newcommand{\TRAFFICGAIN}{2.4$\times$}

\newcommand{\FOOTPRINTGAIN}{14$\times$}

\newcommand{\BATCH}{customized batch-processing}

\newcommand{\HEEWOO}[1]{\textcolor{black}{(Heewoo: #1)}}

\newcommand{\added}[1]{\textcolor{black}{#1}}



\begin{abstract}

\textit{De novo} assembly enables investigations of unknown genomes, paving the way for personalized medicine and disease management.
However, it faces immense computational challenges arising from the excessive data volumes and algorithmic complexity.



While state-of-the-art \textit{de novo} assemblers utilize distributed systems for extreme-scale genome assembly, they demand substantial computational and memory resources.
They also fail to address the inherent challenges of \textit{de novo} assembly, including a large memory footprint, memory-bound behavior, and irregular data patterns stemming from complex, interdependent data structures. 
Given these challenges, \textit{de novo} assembly merits a custom hardware solution, though existing approaches have not fully addressed the limitations.

We propose {\NMP}, a hardware-software co-design that accelerates scalable \textit{de novo} genome assembly through near-memory processing (NMP).
Our channel-level NMP architecture addresses memory bottlenecks while providing sufficient scratchpad space for processing elements.
Customized processing elements maximize parallelism while efficiently handling large data structures that are both dynamic and interdependent.
Software optimizations include customized batch processing to reduce the memory footprint and hybrid CPU-NMP processing to address hardware underutilization caused by irregular data patterns.

\NMP\ conducts the same genome assembly while incurring a \FOOTPRINTGAIN\ smaller memory footprint compared to the state-of-the-art \textit{de novo} assembly.
Moreover, \NMP\ delivers a \NMPGAIN\ performance improvement over the \CPU, with a \TRAFFICGAIN\ reduction in memory operations.
Consequently, \NMP\ achieves \THROUGHPUT\ greater throughput than state-of-the-art \textit{de novo} assembly under the same resource constraints, showcasing its superior computational efficiency.

\end{abstract}



\maketitle

\thispagestyle{fancy}
\fancyhf{} 
\fancyhead[C]{2025 International Symposium on Computer Architecture (ISCA)}
\pagestyle{empty}

\vspace{-1.5em}
\begin{center}
\begin{minipage}{\columnwidth}
\scriptsize
\noindent
© Owner/Author | ACM 2025. This is the author's version of the work. It is posted here for your personal use. Not for redistribution. The definitive Version of Record was published in \textit{ISCA 2025}, \url{https://doi.org/10.1145/3695053.3731056}.
\end{minipage}
\end{center}
\vspace{0.5em}

\section{Introduction}

\textit{De novo} genome assembly is a powerful tool for exploring unknown genome sequences, such as those of new viruses and human microbiomes \cite{ghosh2020pakman,zhou2021ultra}.
Analyzing these uncharacterized genomes is crucial for understanding disease mechanisms\cite{virgin2011metagenomics, sharon2013genomes, donia2014systematic, pereira2019metagenomics}, 
developing novel and personalized medical treatments \cite{pavon2024quetzal, georganas2015hipmer,ghiasi2024megis},
and enabling early detection and management of diseases including cancer, autism, infectious diseases, genetic disorders, type 1 diabetes, autoimmune disorders, and obesity \cite{virgin2011metagenomics, sharon2013genomes, donia2014systematic, gu2023gendp}.
Even with available reference genomes, \textit{de novo} assembly remains essential for identifying rare genomic variants of biomedical interest \cite{zhou2021ultra}.

However, \textit{de novo} genome assembly requires processing massive and exponentially growing genomic datasets.
Individual datasets can reach tens of terabytes, with their runtime memory footprint expanding to 20$\times$ the on-disk size of the input data.
The rapid advancement of DNA sequencing technology and the continuous evolution of viruses and bacteria further accelerate this data growth \cite{zhou2021ultra, ghosh2020pakman, ghiasi2024megis}. 
The volume of sequenced genomic data now exceeds that generated by astronomy, particle physics, and even YouTube, with a doubling rate of approximately every seven months—far outpacing Moore's Law and computing advancements~\cite{zhou2021ultra, gu2023gendp, georganas2018extreme}.


Consequently, \textit{de novo} assembly computation faces significant challenges due to both the sheer volume of sequencing data and the algorithmic complexity (NP-hard) of large-scale genome assembly \cite{georganas2015hipmer, georganas2018extreme}. 
The massive data volume leads to extensive memory footprint requirements~\cite{georganas2018extreme, li2015megahit,zhou2021ultra,angizi2020pim}, and processing this data causes substantial movement overhead from both its volume and low reuse patterns \cite{ghiasi2024megis}.
Additionally, the algorithm complexity of large-scale genome assembly incurs irregular data access patterns \cite{ghosh2020pakman, awan2021accelerating}.
These characteristics make \textit {de novo} assembly process both time- and memory-intensive~\cite{angizi2020pim,zhou2021ultra,ghiasi2024megis,awan2021accelerating},
creating bottlenecks in time-constrained genome analysis tasks \cite{zhou2021ultra}. 
Enabling widespread access to personalized medicine requires processing such vast data volumes rapidly and cost-effectively \cite{ekim2021minimizer}.

While state-of-the-art \textit{de novo} assemblers \cite{georganas2015hipmer,ghosh2020pakman} leverage distributed CPU systems to accelerate assembly, they still require massive computational resources (tens of thousands of CPU cores and hundreds of terabytes of memory) and fail to effectively address the fundamentally memory-bound nature of the assembly process.
Given its challenges, importance, and scale, \textit{de novo} assembly merits a custom hardware solution.
While hardware acceleration approaches \cite{zhou2021ultra,angizi2020pim,awan2021accelerating,wu2024abakus} have emerged but face significant challenges such as discontinued memory technologies \cite{zhou2021ultra}, limited scalability to larger genomes \cite{zhou2021ultra,angizi2020pim,awan2021accelerating}, restricted applicability \cite{awan2021accelerating}, and partial assembly process coverage \cite{wu2024abakus}.

In this work, we present a principled hardware-software co-design solution.
Our key observation reveals that state-of-the-art \textit{de novo} assembly is particularly suited for channel-level near-memory processing, as it addresses memory-bound behavior, reduces data movement overhead, and accommodates large data structures that exceed bank-level processor capacity.

Based on this insight, we present a comprehensive hardware-software solution to handle the excessive memory footprint and irregular data patterns.
Our hardware design features a customized pipelined systolic processing element (PE) as the near-memory processing unit, optimized for managing large, dynamic, and interdependent data structures while maximizing task-level parallelism.
To address dependent operations with irregular access patterns, we implement an integrated inter-PE crossbar switch with customized buffers and inter-DIMM network bridges.

Our software stack complements the hardware through two key optimizations.
First, efficient memory management with customized batch processing reduces the memory footprint.
Second, hybrid CPU-NMP processing balances workloads across processing units, mitigating hardware underutilization caused by irregular data structure sizes and access patterns.






Our evaluation demonstrates that \NMP\ significantly improves memory efficiency and computational performance for \textit{de novo} genome assembly.
Software optimization allows \NMP\ to maintain assembly quality while reducing the memory footprint by \FOOTPRINTGAIN. 
The system achieves a \NMPGAIN\ performance improvement over the \CPU, with \NMPONLYGAIN\ speedup from near-memory processing alone, and an additional \CPUNMPGAIN\ gain from the combined pipelined systolic PE design and CPU-NMP hybrid processing strategy.
Under identical resource constraints, \NMP\ exhibits superior computational efficiency, achieving \THROUGHPUT\ higher throughput compared to \textit{de novo} assembly on a supercomputer. 

Our analysis also exposes a fundamental limitation of GPU-based large-scale \textit{de novo} assembly.
When batch sizes are constrained to fit within current GPU memory capacities (80 GB for H100 and A100) for full human genome assembly, the resulting contig quality (N50) deteriorates by more than 50\%.



The main contributions of this work include:

\begin{itemize}

    \item A systematic analysis demonstrating that state-of-the-art \textit{de novo} assembly algorithms are inherently suitable for near-memory processing acceleration.

    \item A near-memory processing architecture with novel hardware-software co-design that addresses the fundamental challenges of \textit{de novo} assembly through three key innovations:
    (1) near-channel compute offload engines for memory-bound operations,
    (2) efficient memory management with customized batch processing for massive memory requirements, and
    (3) pipelined systolic processing elements integrated with a hybrid CPU-NMP strategy for irregular data patterns from large, dynamic, and interdependent data structures.

    \item An end-to-end system implementation achieving substantial performance gains for large-scale \textit{de novo} assembly: \FOOTPRINTGAIN\ reduction in memory footprint, \NMPGAIN\ speedup over CPU baseline, and \THROUGHPUT\ improvement in throughput, making genome assembly practical for personalized medicine applications.



\end{itemize}

\section{Background} \label{200_background}

\subsection{De Bruijn Graph-based \textit{De Novo} Genome Assembly} \label{210_de_novo}



\begin{figure}[tb]
  \centering
  \includegraphics[width=\linewidth]{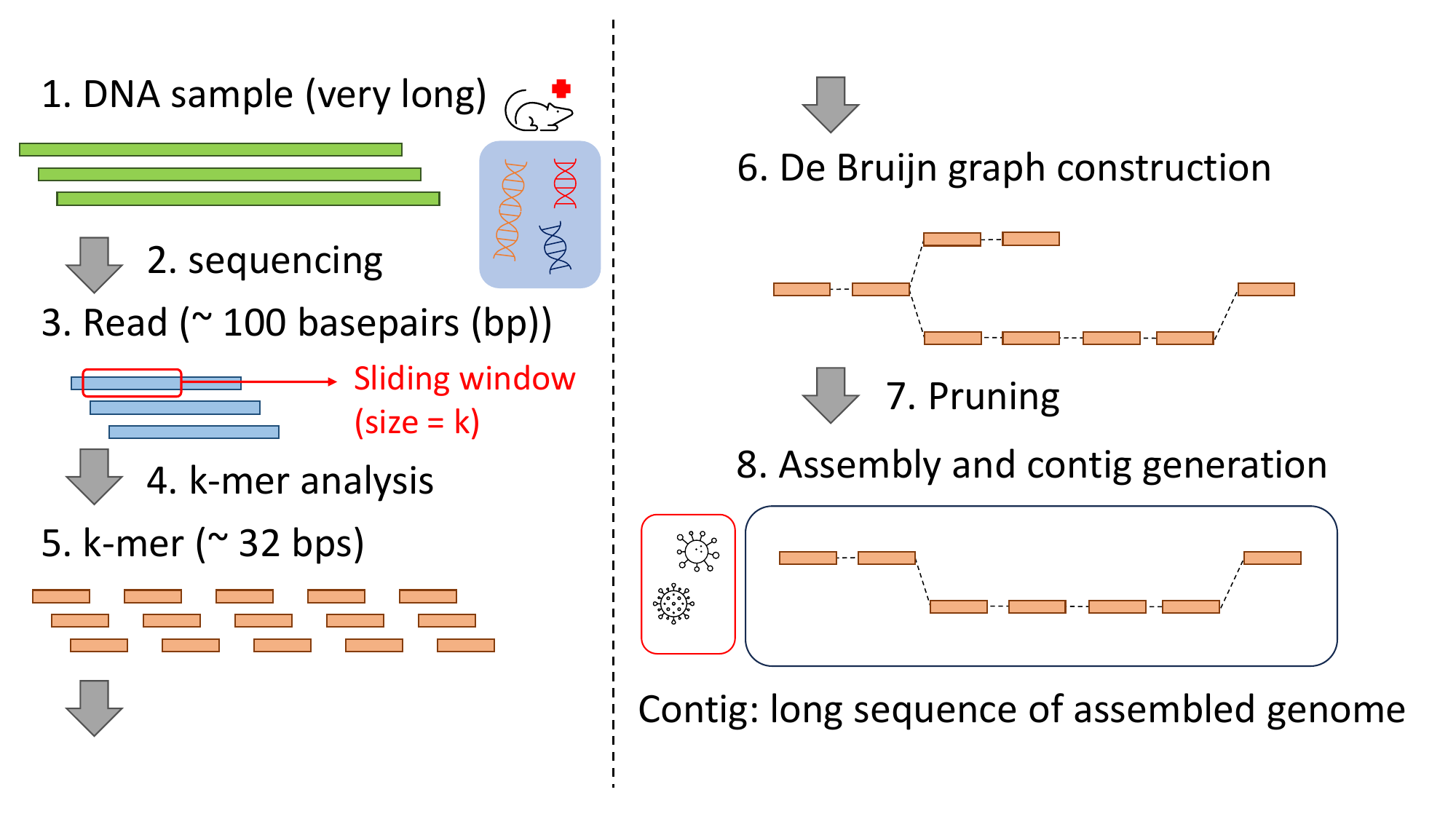}
  \vspace{-8mm}
  \caption{De Bruijn graph-based genome assembly \cite{georganas2018extreme}: 
  It generates contigs (e.g., an unknown virus) from DNA samples. (e.g., an infected mouse).    }
  \label{fig:dbg}
\end{figure}


\textit{De novo} genome assembly reconstructs unknown DNA sequences without the need for a reference genome \cite{ghosh2020pakman}.
For example, when studying an unknown virus, the assembly process generates continuous segments of the target genome (unknown virus) from scratch using DNA samples obtained from an infected host organism \cite{ghosh2020pakman}.

The assembly process varies based on the length of DNA fragments (\emph{reads}) produced by different sequencing technologies: Illumina generates short reads of hundreds of base pairs, while Pacific Bioscience and Oxford Nanopore produce long reads ranging from tens of thousands to hundreds of thousands of base pairs \cite{koren2017canu,ghosh2020pakman}.
While long-read assemblers like CANU \cite{koren2017canu} and hifiasm \cite{cheng2021haplotype} excel at resolving repetitive genome patterns and identifying haplotype information, short-read assembly offers advantages in efficiency, accuracy, cost, throughput, and error rates \cite{ekim2021minimizer,koren2017canu,cheng2021haplotype,ghosh2020pakman}.
This difference is evident in computational requirements: long-read assembly requires 20,000 CPU hours for a human genome assembly~\cite{koren2017canu}, compared to 3,600 CPU hours for short-read assembly \cite{ghosh2020pakman}.
Moreover, short-read sequencing has error rates below 1\%~\cite{ghosh2020pakman}, while long-read error rates of 5-15\%~\cite{cheng2021haplotype} limit their wider adoption~\cite{ghosh2020pakman}.


De Bruijn graph (DBG) assembly serves as the foundational algorithm for short-read \textit{de novo} genome assemblers, including Velvet \cite{zerbino2008velvet}, ALLPATHS \cite{butler2008allpaths}, Abyss \cite{simpson2009abyss}, IDBA-UD \cite{peng2012idba}, Hipmer \cite{georganas2015hipmer}, Megahit \cite{li2015megahit}, and PaKman \cite{ghosh2020pakman}.
A DBG is a directed graph that represents overlapping relationships between \emph{k-mers} (DNA sequences of length k), providing a compact representation of sequence overlap patterns \cite{simpson2009abyss,zhou2021ultra}.

The DBG assembly process iteratively constructs and traverses the k-mer graph to build \emph{contigs} (long contiguous DNA sequences).
As illustrated in Fig. \ref{fig:dbg}, the process begins by extracting k-mers from DNA reads, then uses their overlapping information to construct the graph, and finally traverses the graph to generate contigs.

The DBG genome assembly computation is both time- and memory-intensive because of the substantial amount of sequenced genomes and the excessive number of k-mer nodes in the DBG \cite{zhou2021ultra}.

\subsection{Near-Data Processing}\label{220_NDP}

Near-data processing (NDP) systems move computation into or close to memory and have been widely used for accelerating many AI/ML/graph applications~\cite{zhou2021ultra, ke2020recnmp,li2016pinatubo,feng2022menda,talati2022ndminer}.
NDP is receiving attention from industry leaders, such as Advanced Micro Devices (AMD) and Samsung, as a breakthrough for energy-efficient large-scale computing \cite{su20231}, and it is a publicly available real-world architecture~\cite{gomez2022benchmarking}.
NDP improves the performance of memory-bound workloads by leveraging massive parallelism, minimizing expensive off-chip data transfers from memory to processors (CPU, GPU), and exploiting high internal memory bandwidth for compute units~\cite{talati2022ndminer}.
NDP is also energy-efficient because it reduces the energy-expensive memory-to-processor data transfer, given that data communication consumes 40$\times$ more power than arithmetic operations in processors \cite{fujiki2021near}. 
NDP architectures are implemented in DRAM and other emerging memory technologies \cite{talati2022ndminer}.

NDP systems can be categorized into three groups depending on the proximity of compute units to data \cite{talati2022ndminer}. 

\textbf{(1) Processing-In-Memory:} 
The first group processes data within a memory subarray without reading it out \cite{talati2016logic}, enjoying high internal data bandwidth when operands are aligned in two memory rows/columns \cite{talati2022ndminer,kim2023recpim}. 

\textbf{(2) Bank-Level Near-Memory Processing:}
The second group processes data at the local/global row buffer \cite{lee2021hardware}, which is optimally utilized when operands reside in the same bank \cite{talati2022ndminer}.
Commodity UPMEM PIM-DIMM products adopt this architecture \cite{zhou2023dimm,devaux2019true}.

\textbf{(3) Channel-Level Near-Memory Processing:}
The third group places computation within the buffer chip or memory's logic layers \cite{ke2020recnmp}, allowing access to data from different banks but with limited bandwidth \cite{talati2022ndminer}, as seen in Samsung's AxDIMM \cite{zhou2023dimm,kim2021aquabolt}. 

The placement of compute units within memory depends on the characteristics of the workload \cite{talati2022ndminer}.

\section{Motivation} \label{300_motivation}




\subsection{State-of-the-art \textit{De Novo} Assembly: PaKman}\label{310_pakman}

\begin{figure}[tb]
  \centering
  \includegraphics[width=0.97\linewidth]{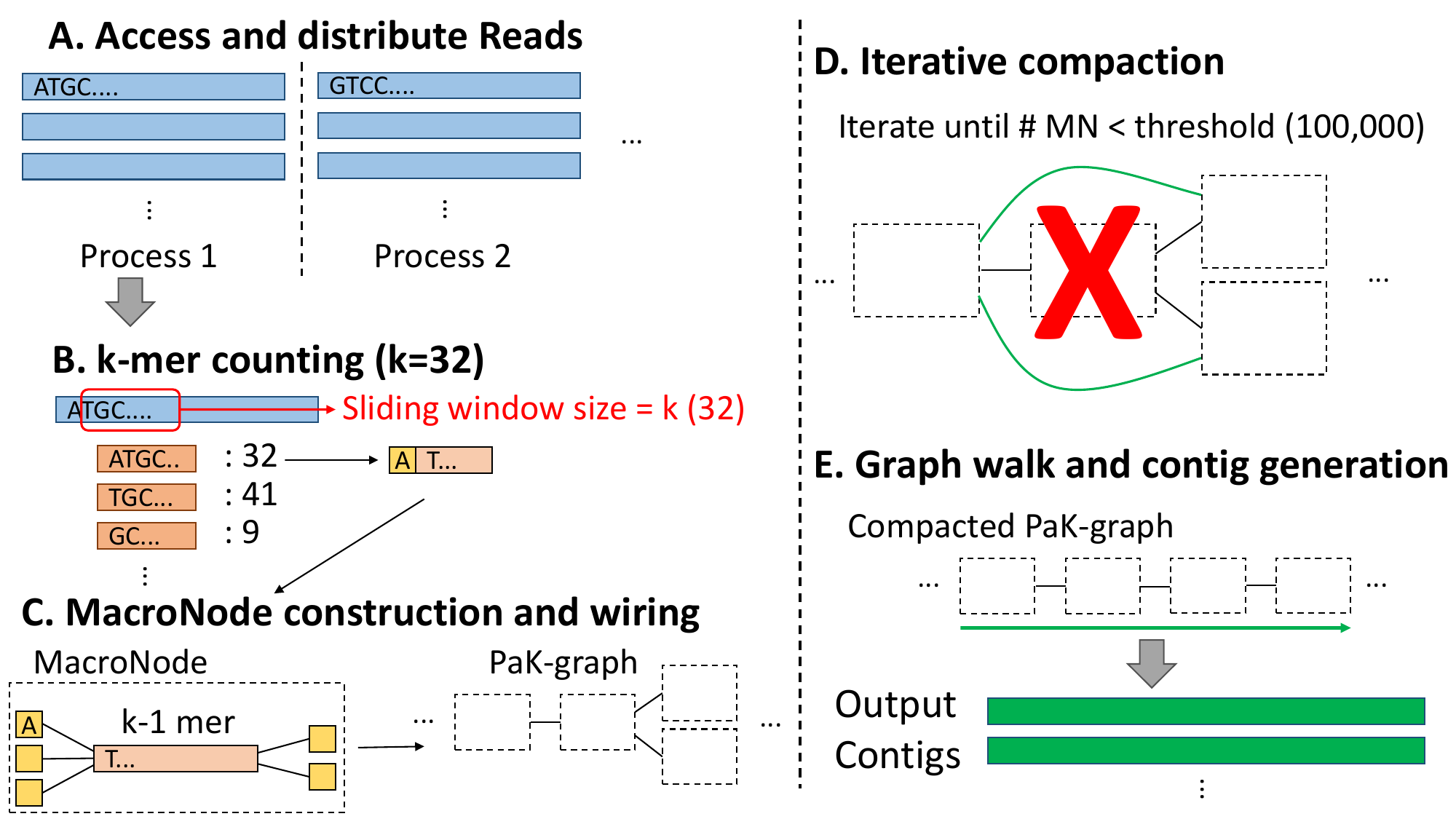}
  \vspace{-3mm}
  \caption{PaKman assembly algorithm procedure \cite{ghosh2020pakman}. 
  A,B: k-mers are generated using a sliding window of size 32. C: MacroNodes are constructed and form the PaK-graphs. D: Iterative Compaction makes the PaK-graphs more compact. E: Contigs are generated from the compacted PaK-graphs.   }
  \label{fig:pakman}
\end{figure}


\begin{figure}[tb]
  \centering
  \includegraphics[width=\linewidth]{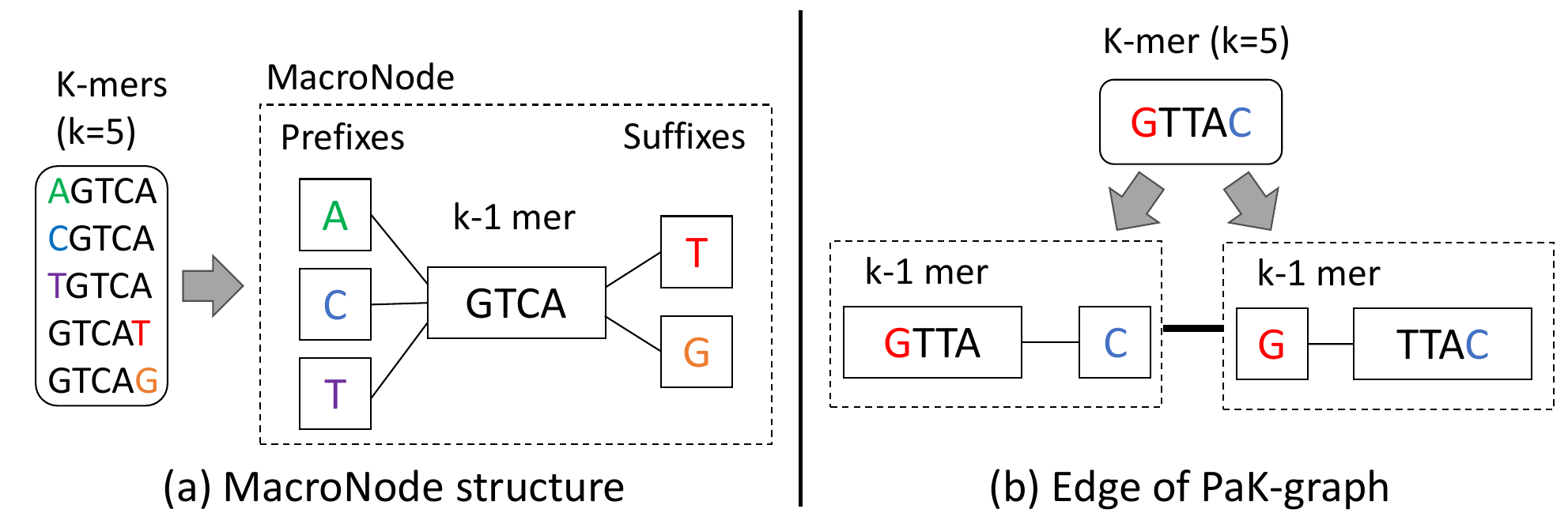}
  \vspace{-7mm}
  \caption{ (a) MacroNode structure and the construction process using k-mers. (b) Creation of PaK-graph edges: one k-mer is used to create two MacroNodes, with that k-mer itself serving as an edge of the PaK-graph. }
  \label{fig:macronode}
\end{figure}

\begin{figure*}[tb]
  \centering
  \includegraphics[width=\linewidth]{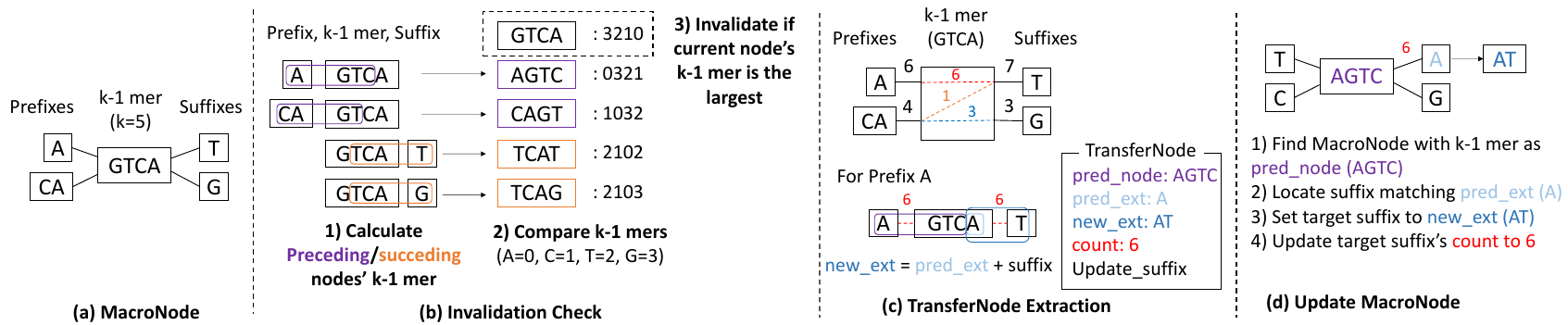}
  \vspace{-5mm}
  \caption{Example of the Iterative Compaction process.
  }
  \vspace{-3mm}
  \label{fig:iterative_compaction}
\end{figure*}


Among the \textit{de novo} assembly algorithms discussed in Section \ref{210_de_novo}, PaKman \cite{ghosh2020pakman} represents the current state-of-the-art approach.
It leverages distributed systems to manage massive memory requirements and enable parallel large-scale genome assembly.
PaKman achieves 3.4$\times$ higher performance than the previous leading assembler, Hipmer \cite{ghosh2020pakman,georganas2015hipmer}, through its innovative data structures and algorithmic strategies optimized for distributed computing environments \cite{ghosh2020pakman}.

Fig. \ref{fig:pakman} shows how PaKman performs the assembly procedure by implementing the de Bruijn graph using \emph{MacroNodes} and \emph{PaK-graphs}.
PaKman's key innovation is the MacroNode data structure (Fig. \ref{fig:macronode}), which groups similar and adjacent k-mers while preserving their interconnections.
This structure enables assembly at the MacroNode level, providing two key advantages: it allows independent and parallel processing of MacroNodes with minimal inter-node communication while also supporting compaction to reduce memory footprint \cite{ghosh2020pakman}.
These MacroNodes form the PaK-Graph, which functions as a distributed De Bruijn graph.

Fig. \ref{fig:macronode} illustrates how MacroNodes are generated and provide a compact data structure while preserving k-mer connectivity and forming the PaK-graph structure.
As shown in Fig. \ref{fig:macronode}(a), k-mers sharing the same (k-1)-mer are grouped into a single MacroNode.
For instance, with k=5, the k-mer ``AGTCA" forms a MacroNode with (k-1)-mer ``GTCA" and prefix `A', while ``GTCAT" forms the same MacroNode with the same (k-1)-mer ``GTCA" but with suffix `T'.
This grouping enables efficient representation where five k-mers can be stored using one (k-1)-mer and five prefixes/suffixes. 

Fig. \ref{fig:macronode}(b) demonstrates how MacroNodes preserve k-mer connectivity.
A single k-mer ``GTTAC" generates two components: one combining prefix `G' with the (k-1)-mer ``TTAC", and another combining the (k-1)-mer ``GTTA" with suffix `C'.
These components are assigned to separate MacroNodes, with edges established between them to maintain the original sequence connectivity.
The MacroNode thus encapsulates k-mer connections through shared (k-1)-mers, while connections between MacroNodes are maintained through overlapping k-mers at their boundaries.

Another key innovation of PaKman is its \emph{Iterative Compaction} process, which optimizes memory usage by progressively merging adjacent MacroNodes \cite{ghosh2020pakman}, as shown in Fig. \ref{fig:pakman} (D).
This reduces the total number of MacroNodes while creating more information-dense structures, ultimately simplifying graph traversal.

During Iterative Compaction, MacroNodes containing the lexicographically largest (k-1)-mer among their neighbors are identified for removal (see Fig. \ref{fig:iterative_compaction} (b)).
Before deletion, their prefix-suffix pairs and internal connectivity information are extracted as \emph{TransferNodes} (see Fig. \ref{fig:iterative_compaction} (c)), which are then transferred to adjacent MacroNodes.
These neighboring MacroNodes update their data structures by incorporating the transferred information (see Fig. \ref{fig:iterative_compaction} (d)), effectively preserving all sequence connections while reducing the total number of nodes.
Fig. \ref{fig:iterative_compaction} illustrates this compaction process.

\begin{tcolorbox}[width=0.48\textwidth]
\textbf{Takeaway 1.}
MacroNode enables parallel processing with minimal inter-node communication.

\textbf{Takeaway 2.}
Iterative Compaction streamlines graph traversal.
\end{tcolorbox}

Although PaKman achieves efficient large-scale genome assembly through its innovative features, it demands extensive computational resources—thousands of CPUs and tens of terabytes of main memory.
These requirements make it impractical for smaller computing environments and cost-prohibitive for individual patient microbiome analysis \cite{virgin2011metagenomics}.

\begin{tcolorbox}[width=0.48\textwidth]
\textbf{Takeaway 3.}
PaKman enables large-scale genome assembly but demands substantial resources.
\end{tcolorbox}

\subsection{New Performance Bottleneck: Iterative Compaction} \label{320_runtime}

\begin{figure}[tb]
  \centering
  \includegraphics[width=\linewidth]{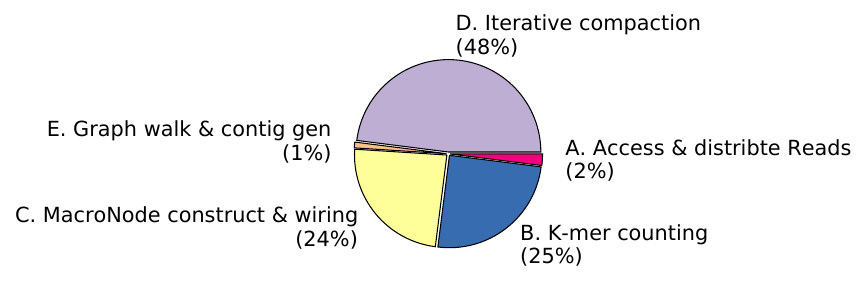}
  \vspace{-7mm}
  \caption{Runtime breakdown of the PaKman algorithm (10\% of the full human genome, 64 threads).
Iterative Compaction dominates the overall assembly time.}
  \label{fig:runtime_proportion}
\end{figure}



While Iterative Compaction effectively reduces graph traversal complexity, it has become the primary performance bottleneck.
As shown in Fig. \ref{fig:runtime_proportion}, the graph traversal step (Step E: Graph walk and contig generation) takes only 1\% of the total runtime, confirming the effectiveness of Iterative Compaction.
However, the same analysis reveals that Step D (Iterative Compaction) itself dominates the assembly process, consuming 48\% of the total runtime.
Further analysis reveals that determining invalid MacroNodes and extracting their prefix, suffix, and wiring information account for 74\% of the compaction time.
These findings highlight the critical need to accelerate the compaction process, particularly the MacroNode invalidation and information extraction stage.

\begin{tcolorbox}[width=0.48\textwidth]
\textbf{Takeaway 4.}
Iterative Compaction dominates the total runtime, emerging as the primary performance bottleneck.
\end{tcolorbox}

\subsection{Memory-Bound Behavior} \label{330_mem_intensive}

\begin{figure}[tb]
  \centering
  \includegraphics[width=0.8\linewidth]{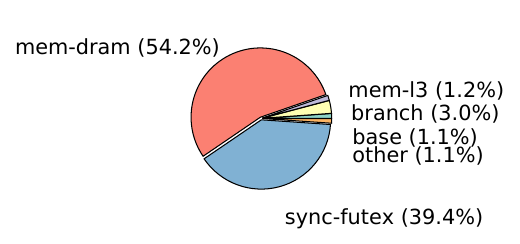}
  \vspace{-5mm}
  \caption{Iterative Compaction stall time breakdown (64 threads): 
  DRAM access (\textit{mem-dram}) dominates, followed by core workload imbalance (\textit{sync-futex}). 
  Other components include core computation (\textit{base}), branch misprediction (\textit{branch}), and last-level cache access (\textit{mem-l3}).  } 
  \label{fig:latency}
\end{figure}


Using the method described in Section \ref{510_implement}, our analysis reveals that DRAM access stalls dominate the runtime.
As shown in Fig. \ref{fig:latency}, these stalls account for 54.2\% of the total execution time, highlighting that Iterative Compaction is fundamentally memory-latency-bound.
Furthermore, memory bandwidth during this step is severely underutilized, achieving only 5.2 GB/s—a mere 2.5\% of the system's 8-channel, 204.8 GB/s capacity.
This reveals untapped parallelism in accessing multiple MacroNodes from memory simultaneously.
These findings highlight the need to address both the memory-latency-bound nature and inefficient bandwidth utilization for improved assembly performance.


\begin{tcolorbox}[width=0.48\textwidth]
\textbf{Takeaway 5.}
Iterative Compaction is memory-latency-bound.

\textbf{Takeaway 6.}
Memory bandwidth is underutilized, highlighting untapped parallelism in simultaneously accessing multiple MacroNodes from memory.


\end{tcolorbox}

\subsection{Irregular Data Access Patterns} \label{340_irregular}

\begin{figure}[tb]
  \centering
  \includegraphics[width=\linewidth]{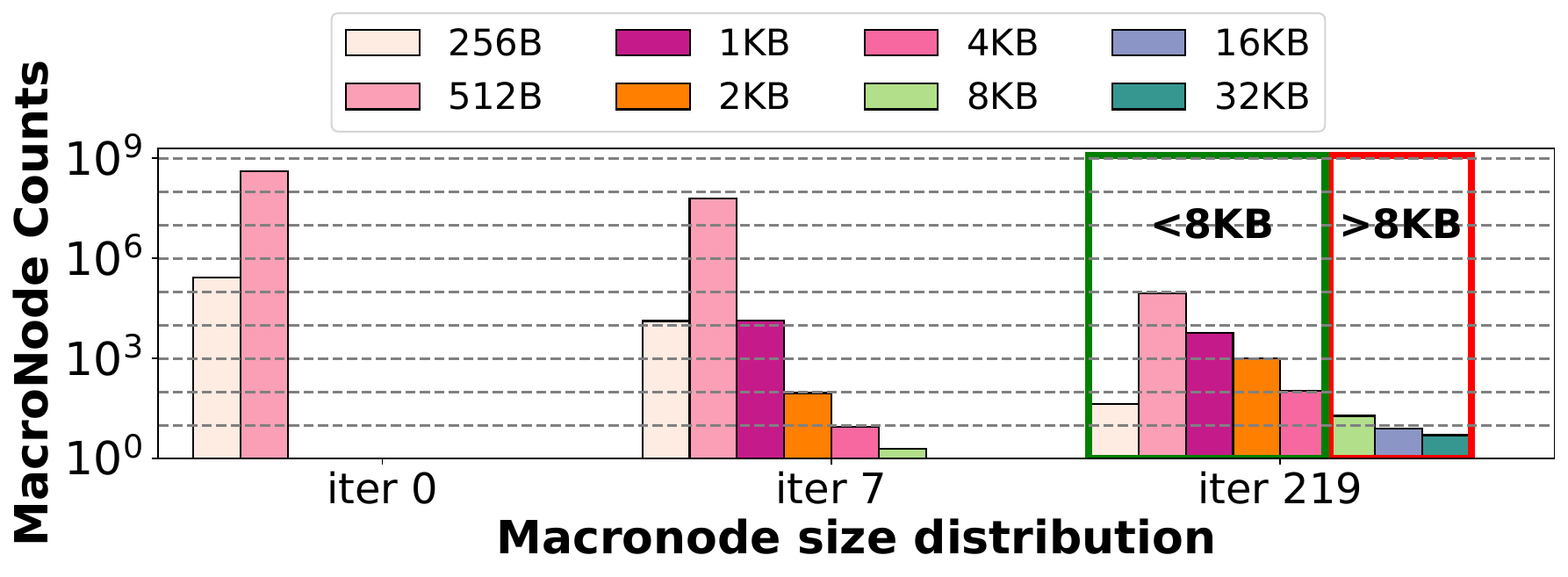}
  \vspace{-5mm}
  \caption{Distribution of MacroNode sizes during Iterative Compaction at iterations 1, 7, and 219 (completion). 
    The y-axis shows the number of MacroNodes in a log scale, and each x-axis label (e.g., 512 B) represents sizes between that value and the next power of two (512 B - 1024 B).
    At completion, only 7.4\%, 1.2\%, 0.1\%, and 0.03\% of MacroNodes exceed 1 KB, 2 KB, 4 KB, and 8 KB, respectively. }  
  \label{fig:mn_distribution}
\end{figure}

\begin{figure}[tb]
  \centering
  \includegraphics[width=\linewidth]{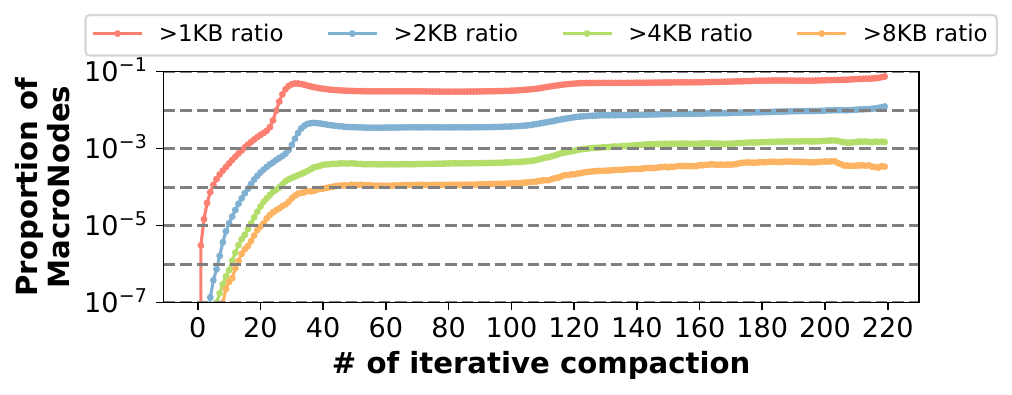}
  \vspace{-7mm}
  \caption{Proportion of MacroNodes exceeding size thresholds during Iterative Compaction.
    Throughout the process, MacroNodes larger than 1 KB, 2 KB, 4 KB, and 8 KB consistently remain below 7.4\%, 1.2\%, 0.16\%, and 0.05\% of total nodes, respectively.
  } 
  \vspace{-5mm}
  \label{fig:mn_8kb_ratio}
\end{figure}



Beyond DRAM stalls, workload imbalance across cores emerges as another significant bottleneck during Iterative Compaction.
As shown in Fig. \ref{fig:latency}, despite distributing equal numbers of MacroNodes across cores, significant \textit{sync-futex} stalls occur due to processing time variations among cores.
This imbalance stems from both the dynamic changes in MacroNode counts and sizes, with the latter following a non-uniform distribution.

MacroNodes are complex data structures comprising multiple 1D and 2D vectors that evolve throughout the Iterative Compaction process.
Their numbers decrease as nodes are invalidated and merged, while their sizes grow as remaining nodes accumulate more k-mer information.
Our analysis of these dynamic changes during Iterative Compaction reveals several key characteristics.

As the process progresses, the distribution of MacroNode sizes becomes wider but shorter (Fig. \ref{fig:mn_distribution}), indicating that while the total number of MacroNodes decreases, certain MacroNodes increase in size due to the growing number and length of prefixes and suffixes.
The size distribution exhibits a long tail, with some MacroNodes expanding up to 32 KB, although the majority still fit within the 8 KB row buffer size.
Further analysis (Fig. \ref{fig:mn_8kb_ratio}) confirms that the proportion of large MacroNodes remains consistently low throughout the entire Iterative Compaction process.
Specifically, MacroNodes exceeding 8 KB constitute less than 0.05\% of the total population across all iterations.

\added{
When the size and count of MacroNodes change, communication occurs between adjacent MacroNodes by extracting TransferNodes from the invalidating MacroNode and transferring them to its neighbors.
The MacroNode graph exhibits complex and irregular connectivity, as reflected in the distribution of adjacent MacroNodes: 8.6\% are connected within the same DIMM, while 91.4\% span across different DIMMs.
This further contributes to irregular memory access patterns and affects the frequency of inter-DIMM communication.
}

The combination of complex MacroNode structures, dynamic size changes, varying node counts during compaction, and the presence of oversized nodes requires a comprehensive solution for efficient processing.  

\begin{tcolorbox}[width=0.48\textwidth]
\textbf{Takeaway 7.}
The dynamic nature of MacroNode sizes and counts leads to irregular data access patterns, resulting in workload imbalance.
\end{tcolorbox}

\subsection{Excessive Memory Footprint} \label{350_footprint}



Despite MacroNode's compact representation of de Bruijn graphs, PaKman still demands substantial memory resources.
Our analysis reveals that during MacroNode construction, wiring, and Iterative Compaction steps, memory usage expands to 13--25$\times$ the original input size.

For example, processing just 10\% of the human genome dataset \cite{human_genome} (38.3 GB) requires 528 GB of memory, while processing 20\% (76.6 GB) exceeds our 1 TB profiling platform capacity.
This memory demand scales proportionally with dataset size, becoming prohibitive for full human genome assembly (383 GB) or patient microbiome analysis (6--12 TB).
Moreover, improving assembly quality through increased sequencing coverage directly impacts memory requirements—a 20\% increase in coverage (from 100 $\times$ to 120 $\times$) results in a corresponding 20\% increase in memory footprint.

These memory requirements significantly limit PaKman's practical applications, particularly in resource-constrained environments or for personalized medicine applications.

\begin{tcolorbox}[width=0.48\textwidth]
\textbf{Takeaway 8.}
Despite their compact nature, MacroNodes still require a substantial memory footprint due to the large number of MacroNodes generated from the extensive volume of k-mers.
\end{tcolorbox}

\section{\NMP\ Design} \label{400_design}


Given these challenges, state-of-the-art \textit{de novo} assembly necessitates a custom hardware solution.
While existing hardware accelerators have not fully addressed these challenges, our systematic analysis of PaKman's performance characteristics reveals the potential of a near-memory processing solution.

In this work, we propose \NMP, a novel architecture that combines software optimization and near-memory processing hardware acceleration to address Iterative Compaction's performance bottleneck while operating with significantly reduced computational and memory resources.

On the hardware side, to address memory-latency-bound behavior \textbf{(Takeaway 5)} and memory bandwidth underutilization \textbf{(Takeaway 6)}, we implement channel-level NMP by placing processing element (PE) arrays within the DIMM buffer chip.
This approach reduces memory operation latency and memory traffic between the memory controller and channel while enabling larger PE buffers to accommodate substantial MacroNode data structures compared to bank-level alternatives.
Additionally, it enables each PE to access its target MacroNode in parallel, allowing multiple MacroNodes to be read simultaneously and enhancing memory bandwidth utilization.


To handle irregular access patterns from complex MacroNode structures \textbf{(Takeaway 7)}, we design pipelined systolic PEs that operate at MacroNode granularity.
This approach provides node-level parallelism through MacroNode-granular processing while enabling intra-node parallelism with overlapped and pipelined operations.
The design manages irregular data dependencies between MacroNodes through both integrated inter-PE crossbar switches with specialized buffers and inter-DIMM network bridges~\cite{zhou2023dimm} for efficient TransferNode routing.

On the software side, to address hardware underutilization caused by workload imbalances from irregular data patterns \textbf{(Takeaway 7)}, we adopt hybrid CPU-NMP processing.
This approach offloads operations of exceptionally large MacroNodes to the CPU, balancing NMP PE performance across MacroNodes while eliminating the need for oversized PE buffers.

To address the large memory footprint \textbf{(Takeaway 8)}, we implement two key optimizations.
First, we introduce customized batch processing, which partitions the input dataset into multiple batches for sequential processing. 
Second, we apply data deduplication through pointer aliasing and indirection, minimizing redundant memory usage by reducing MacroNode copies and movements.





\subsection{Hardware: Channel-Level NMP} \label{410_nmp}

\begin{figure}[tb]
  \centering
  \includegraphics[width=0.9\linewidth]{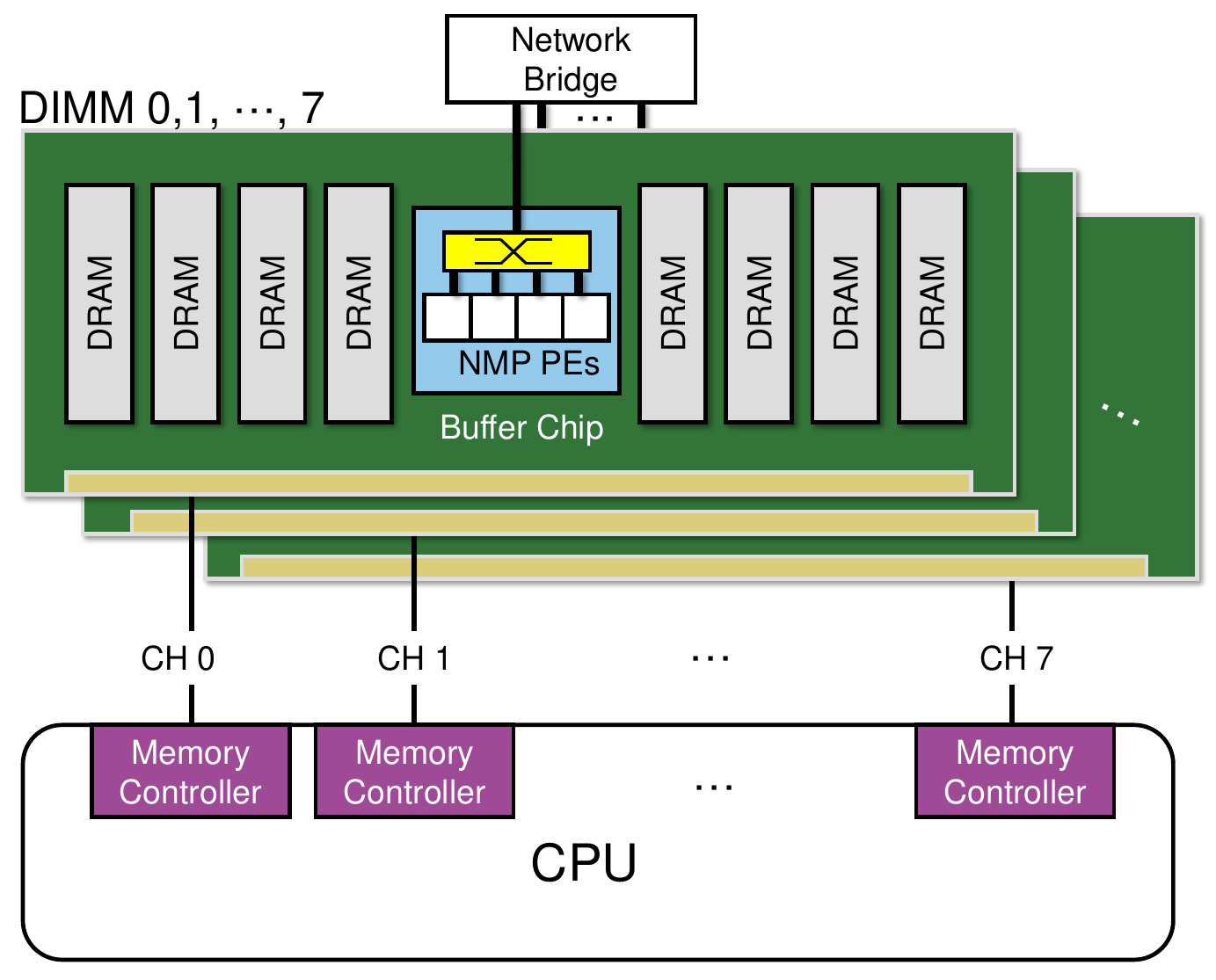}
  \vspace{-5mm}
  \caption{
  \NMP\ architecture: PEs integrated into buffer chips communicate across DIMMs via network bridges.
  } 
  \vspace{-5mm}
  \label{fig:design_overall}
\end{figure}

Iterative Compaction is bounded by memory operations, with a significant portion of runtime spent waiting for data transfer between memory and processing units.
To address this memory-latency bottleneck, we adopt near-memory processing, specifically implementing a channel-level NMP architecture with processing elements (PEs) situated within the DIMM buffer chip.

Our choice of channel-level over bank-level NMP is driven by space requirements: 
substantial PE scratchpad memory for variable-sized MacroNodes (sub KB to several KBs), 
temporary storage for TransferNodes awaiting transfer to neighboring MacroNodes, 
and network switches enabling inter-PE communication between arbitrarily-located neighboring MacroNodes.
The DIMM buffer chip provides sufficient area to accommodate these components. 



As shown in Fig. \ref{fig:design_overall}, our channel-level NMP architecture places PEs within the buffer chip.
\added{The crossbar facilitates the transfer of TransferNodes between different PEs within the same DIMM or across DIMMs via the Network Bridge~\cite{zhou2023dimm}. 
The inputs to the crossbar are connected to each PE's output, which transmits TransferNodes to other PEs, as well as to the Network Bridge output, which receives TransferNodes from PEs in other DIMMs. 
The outputs of the crossbar are connected to each PE’s input, which receives TransferNodes from other PEs, along with the Network Bridge input, which transmits TransferNodes to PEs in other DIMMs.}

\added{For example, in a system with 16 PEs, the crossbar switch would have a 17$\times$17 configuration, where the 16 PE inputs and outputs are connected to the crossbar, along with an input and output port for the Network Bridge.
The Network Bridge transfers TransferNodes between DIMMs using both point-to-point communication and a broadcast mechanism~\cite{zhou2023dimm}.
}


\subsection{Hardware: Pipelined Systolic PE} \label{420_pipeline}

\begin{figure*}[tb]
  \centering
  \includegraphics[width=0.89\linewidth]{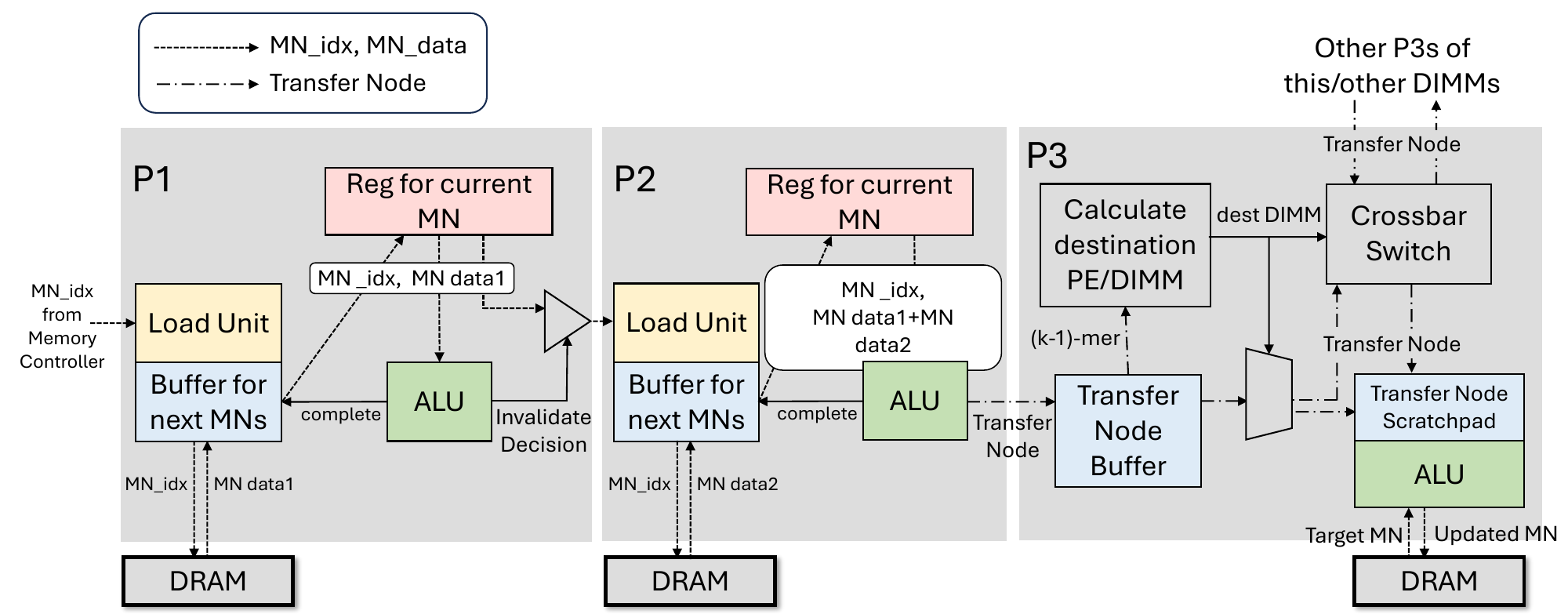}
    \vspace{-3mm}
  \caption{Design of a Processing Element executing Iterative Compaction pipeline stages. 
  \textbf{Stage P1} assesses whether the current MacroNode is a candidate for invalidation. 
  \textbf{Stage P2} extracts TransferNodes from the invalidating MacroNode.
  \textbf{Stage P3} updates the neighboring MacroNode using TransferNode data.
  `MN\_idx' represents the MacroNode's index.
  `MN data1' contains the (k-1)-mer, prefixes and suffixes.
  `MN data2' contains prefix/suffix counts and the internal wiring information.
  }
  \vspace{-3mm}
  \label{fig:design_pipeline}
\end{figure*}



MacroNodes present significant computational challenges due to their large and complex structure, dynamic size changes, and inter-node dependencies through TransferNodes.
Standard NMP systems with general-purpose PE arrays cannot effectively process these complex data structures, making customized PE design essential.

However, the unique properties of MacroNodes also create opportunities for customized acceleration.
While inter-node dependencies exist, most operations occur within individual MacroNodes, enabling node-level parallel processing. 
The compaction process itself comprises a series of separable operations that can be pipelined, providing opportunities for intra-node parallelism.
Furthermore, inter-node communication transfers compact TransferNodes instead of entire MacroNodes, thereby reducing data movement overhead.


Fig. \ref{fig:design_pipeline} illustrates the individual PE design.
Multiple PEs within each buffer chip process MacroNodes in parallel, enabling node-level parallelism.
Each PE implements a 3-stage pipeline for intra-node parallelism.

\textbf{Stage P1 (Invalidation Check)} determines if the current MacroNode is the target for invalidation.
If the MacroNode is the target for invalidation, it advances to Stage P2; otherwise, it terminates processing.
This stage minimizes memory read operations by accessing only the required MacroNode fields: (k-1)-mer, prefixes, and suffixes.

\textbf{Stage P2 (TransferNode Extraction)} extracts TransferNode information from the invalidating MacroNodes received from Stage P1 and forwards these TransferNodes to Stage P3.
This stage optimizes memory access by reusing data from Stage P1 and fetching only additional internal wiring information.

\textbf{Stage P3 (Routing and Update)} determines the destination MacroNode location and manages TransferNode routing.
For destinations within the same PE, TransferNodes are delivered to the local TransferNode Scratchpad.
TransferNodes targeting different PEs within the same DIMM are routed through the crossbar switch.
For destinations in different DIMMs, TransferNodes are routed through the network bridge \cite{zhou2023dimm}.
Finally, Stage P3 updates the destination MacroNode's prefix, suffix, and internal wiring information, and writes the updated MacroNode back to memory.

The computations at each stage consist of simple arithmetic operations (addition and subtraction), shifting, bitwise OR and AND operations, and comparisons.
A key operation across all stages is appending genome base pair sequences, which is implemented using shift and bitwise OR operations.



Stage P1 calculates neighboring MacroNodes' (k-1)-mers by appending the current MacroNode's (k-1)-mer with its prefixes and suffixes.
It then compares the current MacroNode's (k-1)-mer with the calculated neighbors' (k-1)-mers to determine if it is the lexicographically largest, indicating that it is a target for invalidation.
Stage P2 uses appending operations to compute neighboring MacroNodes' (k-1)-mers and their new prefix and suffix extensions.

Stage P3 updates the destination MacroNode's data structure (prefixes, suffixes, and wiring information) using the TransferNode.
For TransferNode destination lookup, it maintains a small mapping table of MacroNode ranges per DIMM, as MacroNodes are stored in ascending (k-1)-mer order across DIMMs, with DIMM 0 containing the lowest (k-1)-mers.
For example, if the mapping table shows DIMM 0's value as 0xff and DIMM 1's value as 0xffff, then MacroNodes with (k-1)-mer values between 0xff and 0xffff are in DIMM 1.
This static mapping eliminates the overhead of searching for MacroNode destinations across DIMMs.

\begin{figure}[tb]
  \centering
  \includegraphics[width=0.9\linewidth]{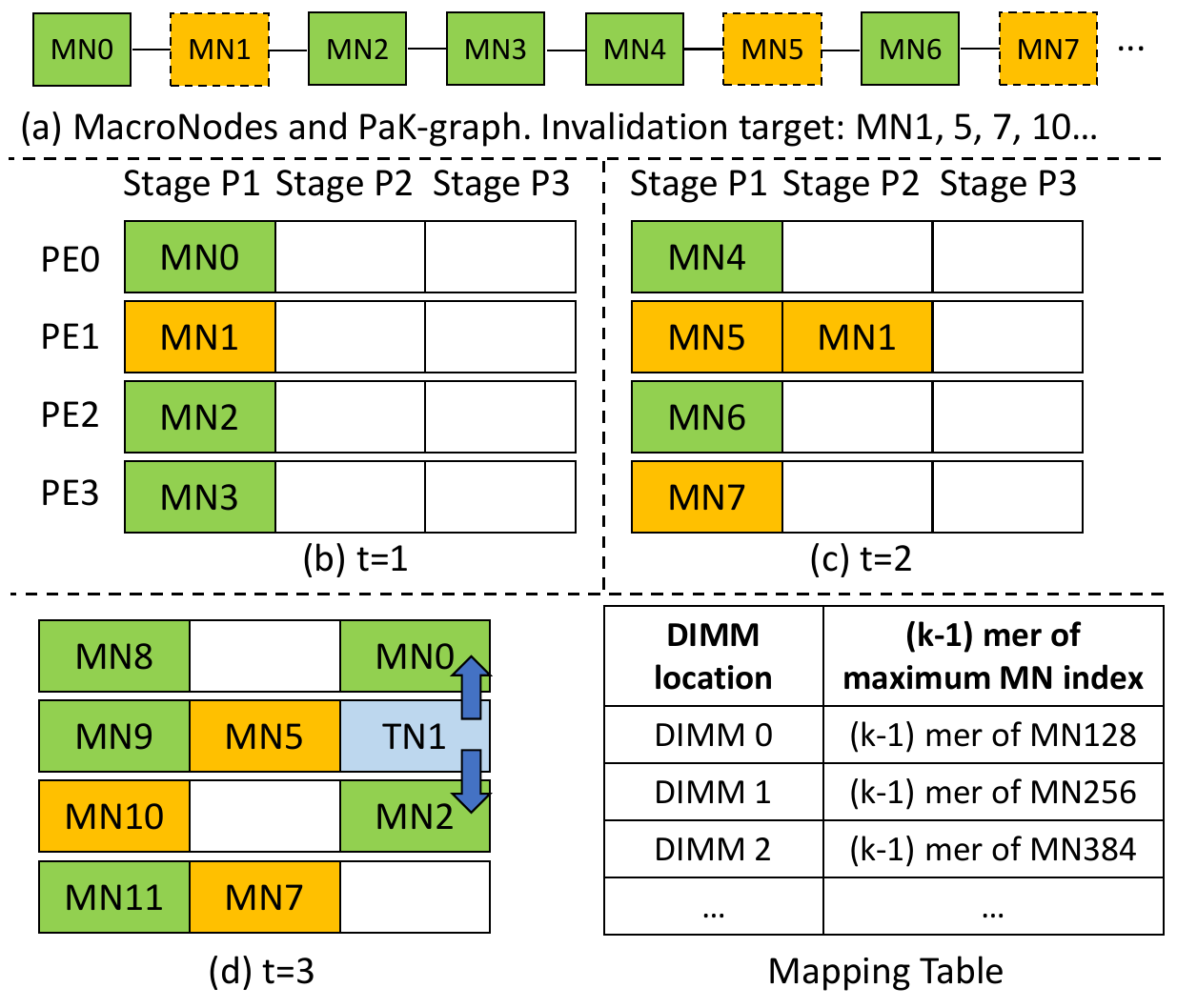}
  \vspace{-5mm}
  \caption{A walkthrough example of pipelined PEs.
  \vspace{-6mm}
  }
  \label{fig:pipeline_example}
\end{figure}

Fig. \ref{fig:pipeline_example} illustrates an example of how pipelined Processing Elements (PEs) perform Iterative Compaction at MacroNode granularity.
Fig. \ref{fig:pipeline_example}(a) shows the input PaK-graph, where MacroNodes (MNs) 1, 5, 7, and 10 are targeted for invalidation.
In Fig. \ref{fig:pipeline_example}(b), during the first cycle, MN0 through MN3 are processed in parallel by PE0 through PE3 at Stage P1.
Fig. \ref{fig:pipeline_example}(c) shows the second cycle, where only MN1 proceeds to Stage P2 for TransferNode extraction.
In Fig. \ref{fig:pipeline_example}(d), during the third cycle, TransferNodes from MN1 (TN1) are sent to PE0 and PE2 to update MN0 and MN2, respectively.
The mapping table shows that TN1's destinations (MN0 and MN2) and source (MN1) are all located in DIMM 0.

\subsection{Software: Hybrid CPU-NMP Processing} \label{430_offload}

Dynamically and non-uniformly changing MacroNode sizes during Iterative Compaction, along with oversized MacroNodes, creates a workload imbalance between processing units, as discussed in Section \ref{340_irregular}.
Large MacroNodes degrade their PE's performance and cause imbalances among PEs in other channels.
Additionally, these oversized MacroNodes require significant PE areas for buffers and scratchpads to store MacroNodes and intermediate results.
Consequently, designing PEs to accommodate all MacroNode sizes becomes impractical.

However, the highly skewed distribution of MacroNode sizes provides an opportunity to address this issue.
Analysis in Section \ref{340_irregular} shows that only a small fraction of MacroNodes are large, while most MacroNodes (92.6\%) range between 256 B and 1 KB, and 99.95\% fit within the row buffer (8 KB), with only a few reaching sizes of tens of kilobytes.


We adopt a hybrid processing strategy that offloads large MacroNode operations to the CPU, optimizing workload balance and PE area overhead.
NMP PEs are designed for smaller MacroNodes while exceptionally large MacroNodes are processed by the CPU.
This approach combines channel-level NMP efficiency for routine cases with CPU resources for outliers.

Due to the small proportion of large MacroNodes, their CPU processing time overlaps with NMP operations.
When offloading MacroNodes larger than 1 KB to the CPU, their computation time is 49.8\% of the NMP computation time for small MacroNodes, enabling effective overlapping without performance degradation.

\added{
Iterative compaction introduces a scenario in which the CPU and NMP components must remain synchronized on an iteration-by-iteration basis. Both the CPU and NMP engines must operate on the same iteration in lockstep to ensure correctness; neither side should move on to the next iteration if the other side is still processing the current one, preventing data races or stale state.
}


\added{Our runtime system makes intelligent decisions about two things: (a) determining whether to process MacroNodes in NMP or on the CPU, and (b) ensuring synchronization between NMP and CPU processing.
The system employs an analytical model (from Section ~\ref{340_irregular}) to determine the processing location based on MacroNode size.
MacroNodes larger than 1 KB are processed by the CPU, while smaller ones are handled by NMP via a command issued to the memory controller.
}
\added{
The runtime system also ensures synchronization by guaranteeing that the CPU and NMP complete all tasks in iteration $i$ before proceeding to $i+1$, preventing premature access to partial results and maintaining correctness in compaction.
}




\subsection{Software: Customized Batch Processing} \label{440_batch}

Near-memory processing with customized PEs addresses memory-latency-bound behavior and irregular data patterns, but it cannot reduce memory footprint and resource requirements.
Given the substantial size of the input dataset and the even larger resulting MacroNode sizes, it is impractical to process all DNA reads simultaneously on a single-node computer.

We divide the input dataset into multiple batches, processing each sequentially through Iterative Compaction.
The compacted PaK-graphs from all batches are merged for contig generation.
Due to the small size (tens of MB) of compacted PaK-graphs, merging and traversal incur minimal overhead.
\textbf{Using this approach with another software optimization described in Section \ref{450_sw_opt} and a batch-size of 10\% of the input genome reduces the memory footprint by \FOOTPRINTGAIN\ compared to the original PaKman algorithm.} 

However, reducing the batch size involves a trade-off: while it lowers memory usage, a batch size that is too small adversely affects the quality of the generated contigs.
We employ the N50 metric to evaluate contig quality and analyze the relationship between batch size and contig quality. 
N50 is a widely recognized metric used to assess the contiguity of an assembly \cite{gurevich2013quast}, representing the length of the smallest contig such that contigs of this length or longer cover at least 50\% of the total assembly. 
Consequently, a larger N50 value indicates less fragmentation and thus higher quality of the contigs, assuming the total assembly size remains constant. 
As indicated in Table \ref{tab:n50}, smaller batches tend to yield lower contig quality metrics. 
However, when the batch size is approximately 5\%, the quality becomes comparable to that achieved by the distributed system. 
It is important to note that while N50 provides a useful approximation, it is a simplistic indicator of contig quality.

\begin{table}[]
\centering
\resizebox{\columnwidth}{!}{
    \begin{tabular}{|c|c|c|c|c|c|c|c|}
    \hline
    \makecell{Genome \\batch size} & 0.5\% & 1\%  & 3\%  & \makecell{4\% \\ (GPU max)} & 5\% & \makecell{10\% \\ (\NMP)} \\ \hline
    N50              & 875   & 1123 & 1209 & 1107 & 3014 & 3535                             \\ \hline
    \end{tabular}
}
\caption{Contig quality across various batch sizes.}
\vspace{-10mm}
\label{tab:n50}
\end{table}



\subsection{Software: Refined PaKman Algorithm} \label{450_sw_opt}




We implement software-level optimizations to improve performance by exploiting additional parallelism, decrease memory footprint via efficient memory management, and reduce memory operations through a modified Iterative Compaction.

\paragraph{\textbf{Improved Parallelism.}}
Even with only 10\% of the human genome dataset, the initial assembly process on a single node using 64 threads took 26.75 hours, dominated by k-mer counting (25.41 hours) and Iterative Compaction (1.36 hours).
Analysis revealed parallelizable sections in both phases.


\added{Prior work's~\cite{ghosh2020pakman} k-mer counting method has primarily focused on multi-node computing using MPI, aiming to reduce the computational load per node.
However, it underutilizes parallelism within individual nodes.
Additionally, it does not assume large-scale data handling per each node, leading to inefficient memory allocation, resulting in increased runtime.
To address these issues, we introduce three optimizations:
}

\added{
\textbf{(a) Parallel sliding window for k-mer extraction.}
Given that input Read sequences have a fixed length of 100 base pairs, we can precompute their starting addresses with negligible overhead.
Using these addresses, we initialize multiple pointers at the start of each Read and apply a parallel sliding window approach with OpenMP.
}

\added{
\textbf{(b) Preallocating vector space for k-mers.}
Prior method sequentially stores extracted k-mers by pushing them into a single vector without preallocating memory, causing frequent reallocations.
Since \texttt{std::vector} expands exponentially (doubling capacity each time it exceeds its limit), this approach becomes highly inefficient when handling tens of billions of k-mers (hundreds of GBs).
As an improvement we allocate separate vectors per each thread.
With 64 threads, each vector is significantly smaller.
When merging these vectors, we precompute the total required size and preallocate memory for the final vector, which minimizes costly memory reallocations during k-mer storage.
}

\added{
\textbf{(c) Parallel k-mer sorting.}
To count duplicate k-mers, the extracted k-mers are sorted lexicographically.
Instead of the serial sorting approach used in prior work, we utilize parallel sorting (\texttt{\_\_gnu\_parallel::sort}), significantly accelerating the process.
}


In Iterative Compaction, we apply MacroNode-level parallelization when determining invalidated targets and updating MacroNodes using TransferNodes.
While determining invalidated targets is fully parallelizable due to no dependencies, we use \texttt{omp\_set\_lock} for atomic operations when multiple TransferNodes update the same MacroNode.

These optimizations yield substantial performance improvements. 
The total assembly time decreased from 26.75 hours to 0.24 hours (110$\times$ faster), with k-mer counting time reducing from 25.41 hours to 0.06 hours (416$\times$ faster) and Iterative Compaction time improving from 1.36 hours to 0.12 hours (11.6$\times$ faster).
This parallel implementation serves as the baseline for further acceleration, and all profiling in Section \ref{300_motivation} uses this improved algorithm.

\paragraph{\textbf{Efficient Memory Management.}}



\added{
Prior work~\cite{ghosh2020pakman} exhibits inefficient memory usage by storing \texttt{MacroNode} structs directly in \texttt{MN\_map}, a mapping from (k-1)-mers to \texttt{MacroNode} values.
Since \texttt{MacroNode} structs are passed by value in function calls, each invocation redundantly allocates memory and duplicate data in the call stack.
To eliminate this overhead, we modify \texttt{MN\_map} to store \texttt{MacroNode} pointers instead, ensuring that function calls operate on references rather than creating unnecessary copies, as shown below:
}
\verb|void func(vector<pair<kmer_t,MacroNode*>> &MN_map){..}|.

\added{
This refactoring eliminates the creation of duplicate \texttt{MacroNode} copies at each function call, reducing both memory allocation overhead and stack memory usage. 
}

We also postpone the deletion of invalidated MacroNodes until after Iterative Compaction completes, reducing MacroNode movement and reallocation overhead.
These optimizations reduce peak memory requirements by 1.4$\times$ for the 10\% human dataset, from 528 GB to 379 GB.

\paragraph{\textbf{Optimize Process Flow for Less Memory Operations.}}
We modify Iterative Compaction to better suit NMP acceleration by restructuring its execution order and step granularity.
The original algorithm processes each stage of Iterative Compaction sequentially, requiring all MacroNodes to complete their current stage before any MacroNode can proceed to the next stage.
This sequential approach also incurs substantial memory operation overhead as each stage must access all MacroNodes before proceeding, resulting in repeated memory operations across the entire MacroNode set.

Our modified design processes MacroNodes individually in a pipelined systolic manner, allowing MacroNodes to advance to subsequent stages independently.
This means one MacroNode can proceed to the next stage while others are still processing previous stages or waiting to be issued, overlapping the computation time.
This approach also enables data reuse between stages, as data transferred from previous stages can be immediately utilized by subsequent stages, thereby reducing both memory operations and traffic.

\subsection{Exploring Alternative Designs} \label{460_explore}



\paragraph{\textbf{Hybrid GPU-CPU architecture with NMP}}

\added{This architecture can offload k-mer counting to the GPU, which is highly parallelizable and accounts for 25\% of the assembly time. 
However, challenges such as significant GPU-to-{\NMP} data transfers involving large volumes of data (333 GB for a 10\% human batch) and potential coherence issues remain, requiring further investigation.}

\paragraph{\textbf{Near-storage computing}}
\added{
It reduces data movement overhead but faces challenges such as unnecessary page reads due to irregular, fine-grained data patterns, SSD wear-out from frequent writes during iterative compaction, and limited data read bandwidth (7 GB/s~\cite{ssd_bw, xu2024performance}) compared to {\NMP} (25.6 GB/s for DIMM reads and 25 GB/s for inter-DIMM communication~\cite{zhou2023dimm}).
}

\paragraph{\textbf{Extending {\NMP} for general-purpose acceleration.}}
\added{
{\NMP} utilizes massive internal bandwidth, supports parallel integer arithmetic, and enables inter-PE communication.
However, it lacks support for floating-point arithmetic, matrix operations, and programmable dataflow, which could enable broader workload flexibility but comes at the cost of increased area overhead and reduced efficiency for \textit{de novo} assembly.
}




\section{Methodology} \label{500_method}

\begin{table}[]
\centering
\begin{tabular}{c}
\hline \hline
\textbf{DNA sequencing}  \\ \hline
ART simulation tool \cite{huang2012art}, coverage=100$\times$ \\
read length=100, k-mer size=32  \\
Full human genome\cite{human_genome}, batch-size=10\% \\ \hline \hline
\textbf{PaKman profiling platform /Host CPU (CPU baseline)} \\ \hline
\begin{tabular}[c]{@{}c@{}}Intel Xeon Platinum 8380 CPU, 2.3 GHz, 40 cores (80 threads)\\ DDR4-3200 MT/s, 1 TB, 8-channels, 2 ranks per channel, \end{tabular}       \\ \hline \hline
\textbf{PaKman profiling tools}  \\ \hline
Linux perf, Sniper simulator \cite{carlson2014aeohmcm}  \\ \hline \hline
\textbf{NMP Implementation} \\ \hline
\begin{tabular}[c]{@{}c@{}}Ramulator simulator \cite{kim2015ramulator}\\ DDR4-3200 MT/s, 1 TB, 8-channels, 2 ranks per channel\\ PE frequency @ 1.6 GHz \\ Buffer sizes: MacroNode (4 KB), TransferNode (1 KB) \end{tabular} 
\\ \hline \hline
\end{tabular}
\caption{Implementation methods and system parameters.}
    \vspace{-8mm}
\label{tab:method}
\end{table}

\subsection{Implementation Methods and System Parameters}\label{510_implement}

The implementation methodology and system parameters are outlined in Table \ref{tab:method}.
We use the ART simulation tool \cite{huang2012art} to sequence the sample DNA into reads.
The reads are sequenced as 100 base pairs each, with a coverage of 100$\times$.
We use the full human genome \cite{human_genome} to provide a comparison with previous PaKman work, employing a 10\% batch size to minimize the memory footprint without sacrificing contig quality.
Memory bandwidth profiling is conducted using Linux perf on two Intel Xeon Platinum 8380 CPUs, each featuring 2.3 GHz with 40 cores (80 threads).
The system has 1 TB of memory with an 8-channel DDR4-3200 MT/s configuration.
Stall times are profiled using the Sniper simulator \cite{carlson2014aeohmcm}.

The memory configuration for the NMP system comprises DDR4-3200 MT/s with 8 channels and 2 ranks per channel, totaling 1 TB in capacity, while the PE operates at a frequency of 1.6 GHz.
Each MacroNode buffer size is 4 KB, and each TransferNode buffer/scratchpad size is 1 KB. 
MacroNodes larger than 1 KB are offloaded to the CPU. 

\subsection{Simulation Methodology}
For the implementation and evaluation of our NMP system, we implement a cycle-accurate simulator using Ramulator~\cite{kim2015ramulator}.
As a widely validated DRAM simulator that has been tested against actual DRAM hardware, Ramulator is used in numerous studies \cite{talati2022ndminer,ke2020recnmp,yan2019alleviating} to evaluate NMP system performance compared to CPU baselines.
We faithfully model PEs within Ramulator and their execution time based on the RTL design and the instruction count statistics for each stage.
Additionally, we incorporate the crossbar network delay into the TransferNode routing process.

We generate memory traces of read and write operations from the actual assembly execution to feed them into Ramulator. 
Since MacroNode size exceeds the LLC block size (64 B), a single MacroNode read/write operation generates multiple traces.
To group those traces for each MacroNode, we use `mn\_idx' metadata to control their operation timing and track their status.
When a MacroNode is issued, all traces with the corresponding `mn\_idx' are issued simultaneously.
After completing the MacroNode read, the pipeline stage waits for the modeled execution cycles before sending the MacroNode to the next stage.

\subsection{Baselines} \label{530_baselines}

Below are the configurations of the baselines and our NMP system as detailed in our study:

\textbf{\CPU:}
  This baseline represents a software-optimized PaKman operation with a reduced memory footprint, as described in Section \ref{440_batch} and Section \ref{450_sw_opt}. 
  In this setup, data is transferred from memory to the CPU for processing and then written back to memory. 
  As discussed in Section \ref{450_sw_opt}, each step of Iterative Compaction is processed sequentially.
  The detailed configuration of this baseline is provided in Table \ref{tab:method}.

\added{\textbf{GPU baseline:}
This baseline models an NVIDIA A100 40GB GPU, which is comparable in cost to an Intel Xeon 8380.
A subset of traces with a memory footprint of less than 40GB is used to evaluate the GPU baseline against others.
We simulate the GPU memory system using parameters similar to those of the A100.
} 



\textbf{\NMP\ (Proposed solution):}
  This NMP system incorporates the optimizations discussed in Section \ref{400_design}, beginning with 32 Processing Elements (PEs) per channel.
  The main memory configuration of this system, shown in Table \ref{tab:method}, is the same as that of the \CPU, with the PEs operating at a frequency of 1.6 GHz. 
  
\textbf{\CPUNMP:}
  This configuration retains the optimizations from Section \ref{400_design} but executes computations on the CPU rather than within the NMP PEs.
This setup evaluates performance improvements solely achieved from near-memory processing.
It uses the same configuration as the \CPU.
  
   

\textbf{\NMPPE:}
  This design features infinitely fast Processing Elements (PEs), enabling each stage to compute in a single cycle. 
  Consequently, the runtime is determined solely by the data read and write operations.
  This system aims to evaluate whether adopting faster PEs can further enhance performance and whether \NMP\ fully utilizes memory bandwidth.
  
\textbf{\NMPFW:}
  This configuration includes forwarding logic between \textbf{Stage P1} and \textbf{Stage P3} in the pipelined PEs described in Fig. \ref{fig:design_pipeline}.
  We adopt ideal forwarding logic, which ensures that all MacroNodes data read in \textbf{Stage P1} is reused fully in \textbf{Stage P3}, thereby eliminating duplicated memory read operations. 
  This approach allows us to determine the theoretical maximum performance achievable through reducing memory traffic and maximizing data reuse.




\section{Evaluation Results} \label{600_evaluation}

\subsection{Performance Analysis} \label{610_performance}

\begin{figure}[tb]
  \centering
  \includegraphics[width=\linewidth]{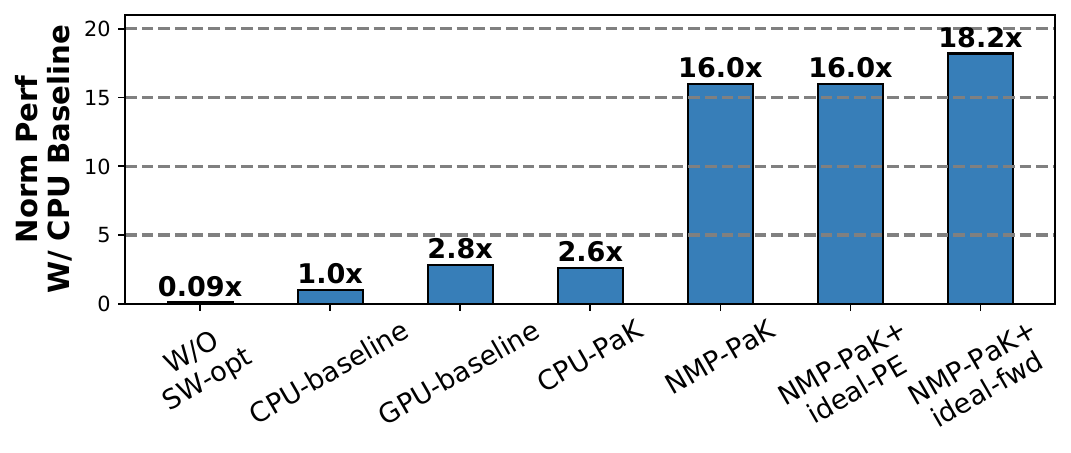}
  \vspace{-8mm}
  \caption{
  Performance improvement of \NMP. 
  All values are normalized to the \CPU.
    \emph{W/O SW-opt} represents the performance prior to the parallelism improvements described in Section \ref{450_sw_opt}.  
  }  
  \label{fig:performance}
\end{figure}



\textbf{\NMP\ vs. \CPU.}
Fig. \ref{fig:performance} illustrates the comparative performance of the \NMP\ against the \CPU. 
The results demonstrate that the \NMP\ outperforms the CPU baseline by a factor of \NMPGAIN. 
Such significant speedups are attributed to several key design optimizations: 
(a) Channel-level NMP, which addresses the memory-latency bottleneck;
(b) Pipelined Iterative Compaction, which maximizes parallelism while handling data dependencies;
and (c) Software optimizations that balance workloads across non-uniform MacroNode sizes and optimize process flow. 


\textbf{\NMP\ vs. \CPUNMP.}
The performance benefit derived from near-memory processing alone can be evaluated by comparing \NMP\ and \CPUNMP. 
Fig. \ref{fig:performance} shows a remarkable performance improvement of \NMP\ over \CPUNMP\ by \NMPONLYGAIN\ (\NMPGAIN\ vs. \CPUNMPGAIN), underscoring the specific advantages of near-memory processing in accelerating memory-latency-bound applications.

\textbf{\CPUNMP\ vs. \CPU.}
The performance gain from our optimization techniques, aside from near-memory processing, is indicated in \CPUNMP. 
As shown in Fig. \ref{fig:performance}, \CPUNMP\ shows a performance increase of \CPUNMPGAIN\ over the \CPU, highlighting the effectiveness of our optimization techniques independent of near-memory processing. 

\added{\textbf{{\NMP} vs. GPU baseline.}
While the GPU baseline achieves a 2.8$\times$ speedup over the {\CPU} (Figure \ref{fig:performance}) due to its massive parallelism and high memory bandwidth, it still significantly underperforms relative to {\NMP} by 5.7$\times$.
This performance gap arises because {\NMP} effectively handles fine-grained, irregular memory access patterns by leveraging channel-level NMP with pipelined systolic PEs.
These findings highlight the computational efficiency of {\NMP} for \textit{de novo} assembly, demonstrating its practicality and affordability.}

\begin{figure}[tb]
  \centering
  \includegraphics[width=\linewidth]{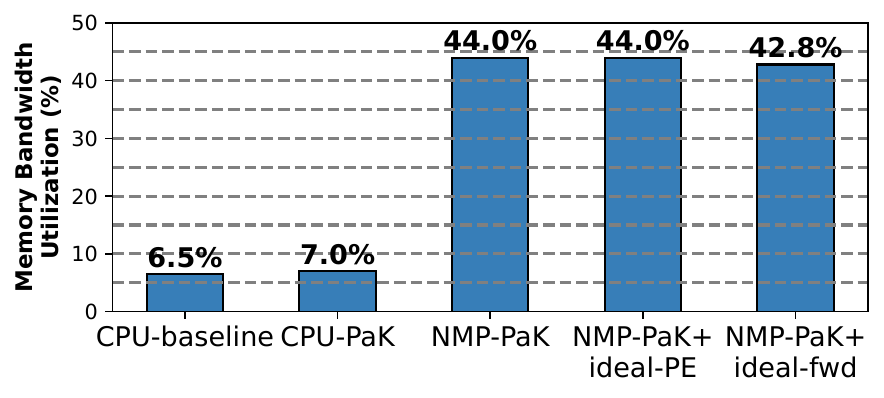}
  \vspace{-8mm}
  \caption{Memory bandwidth utilization.
  \NMP\ utilizes memory bandwidth better than the \CPU. }  
  \vspace{-3mm}
  \label{fig:mem_bw}
\end{figure}

\textbf{Memory bandwidth utilization.}
As shown in Fig. \ref{fig:mem_bw}, \NMP\ achieves significantly higher memory bandwidth utilization (44\%) compared to the \CPU\ and \CPUNMP\ (6.5--7 \%).
Although {\CPUNMP} shows a slight improvement, it is not sufficient to address memory bandwidth underutilization.
These results emphasize the benefit of near-memory processing, which significantly improves parallelism in accessing multiple MacroNodes from memory simultaneously.

\begin{figure}[tb]
  \centering
  \includegraphics[width=\linewidth]{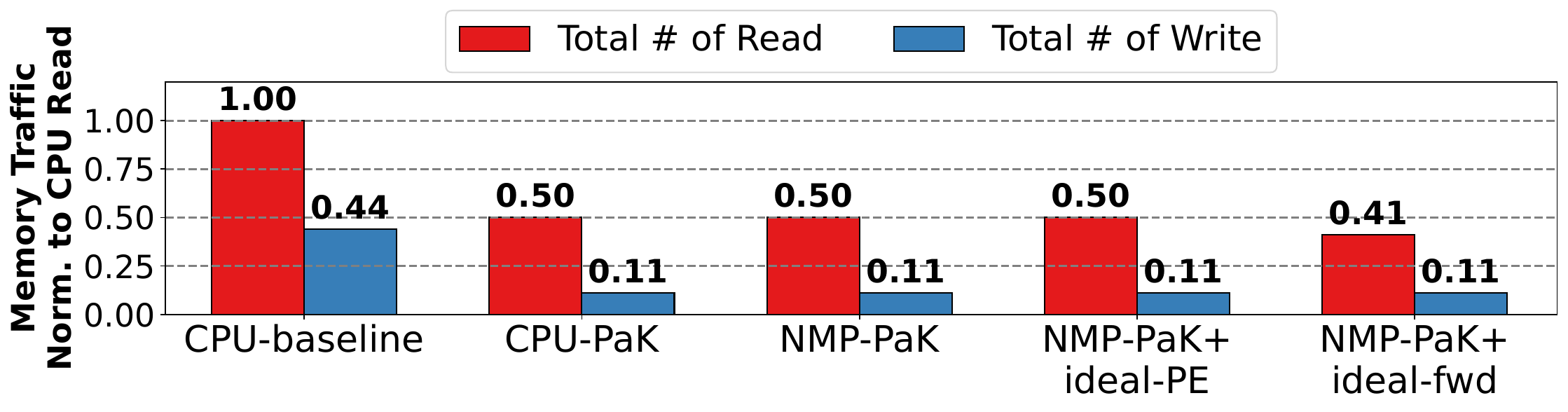}
  \vspace{-7mm}
  \caption{Read and write memory traffic comparison.
  All values are normalized to the {\CPU} read traffic. 
  {\NMP} exhibits lower memory traffic than the \CPU. }  
  \vspace{-5mm}
  \label{fig:mem_traffic}
\end{figure}

\textbf{Memory traffic reduction.}
Fig. \ref{fig:mem_traffic} shows that \NMP\ and \CPUNMP\ have overall \TRAFFICGAIN\ fewer memory operations compared to the \CPU.
This includes 2$\times$ fewer read operations (1.00 to 0.50) and 4$\times$ fewer write operations (0.44 to 0.11). 
This result illustrates the advantages of pipelined Iterative Compaction and the optimized process flow, which increase data reuse between each stage, thereby reducing the number of memory read and write operations and alleviating the memory traffic.


\textbf{\NMP\ with ideal PE.}
We evaluate \NMPPE\ to determine whether adopting faster and more powerful PEs is necessary and to assess if the memory bandwidth is fully utilized.
As shown in Fig. \ref{fig:performance} and \ref{fig:mem_bw}, the performance and memory bandwidth utilization of \NMPPE\ are the same as those of \NMP.
These results indicate that the PEs are not a performance bottleneck and affirm the competency of \NMP\'s PE design. 
Additionally, this confirms that \NMP\ fully utilizes the memory bandwidth.

\textbf{\NMP\ with ideal forwarding logic.}
We test \NMPFW\ to determine the upper limit of performance achievable by maximizing data reuse and minimizing memory traffic.
As shown in Fig. \ref{fig:performance} and \ref{fig:mem_traffic}, the ideal forwarding logic provides a {\FWGAIN} performance boost (\NMPGAIN\ to \NMPFWGAIN) and an \FWTRAFFIC\ reduction in read operations (0.50 to 0.41).
The marginal performance improvement of \NMPFW\ indicates that \NMP\ already handles data reuse and memory traffic reduction effectively, approaching the theoretical maximum.

While implementing elaborate forwarding logic could bring performance closer to this upper limit, it requires significant design cost and complexity.
This includes not only the forwarding logic hardware but also an advanced data allocation and mapping scheme to maximize the temporal and spatial locality of the MacroNodes processed simultaneously in \textbf{Stage P1} and \textbf{Stage P3} of the pipeline. 

To conclude, \NMP\ represents an optimal design in terms of performance and design cost.

\subsection{Sensitivity Study} \label{620_sensitivity}

\begin{figure}[tb]
  \centering
  \includegraphics[width=\linewidth]{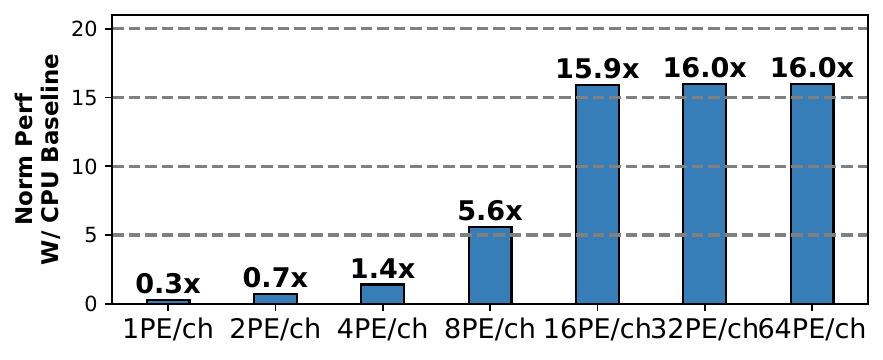}
    \vspace{-7mm}
  \caption{Comparison of \NMP\ performance across various PE/channel. 
  Performance improves as the number of PEs/ch increases, reaching saturation at 32 PEs/ch.}
    \vspace{-5mm}
  \label{fig:pe_sensitivity}
\end{figure}

We examine the impact of varying the number of PEs per channel on \NMP\ performance. 
As depicted in Fig. \ref{fig:pe_sensitivity}, performance initially increases with more PEs per channel but plateaus at 32, remaining unchanged at 64. 
While the performance difference between 16 and 32 PEs per channel is minimal, the latter incurs a 2$\times$ area overhead. 
Therefore, 16 PEs per channel offers a more cost-effective balance between performance and area efficiency.

\subsection{Proportion of intra- and inter-DIMM Communication} \label{625_network}

\added{Intra-DIMM communication accounts for 12.5\%, while inter-DIMM communication comprises 87.5\% of total communication.
Among intra-DIMM communication, 6\% occurs within the same PE, while 94\% is directed to a different PE (in the 16-PE case), highlighting the irregular connectivity between MacroNodes and demonstrating the effectiveness of the crossbar switch and Network Bridge.}
\subsection{Comparison with a Supercomputer} \label{630_distributed}

We evaluate the efficiency of \NMP\ by comparing it with the performance of PaKman on a supercomputer, as reported by Ghosh et al. \cite{ghosh2020pakman}. 
According to our findings from Sections \ref{450_sw_opt} and \ref{610_performance}, the assembly time for the 100\% full human genome on a single node \NMP\ is 4,813 seconds. 
In contrast, the PaKman distributed system completed the assembly in 39 seconds using 16,384 cores across 1,024 nodes \cite{ghosh2020pakman}.
This results in the supercomputer being 123$\times$ faster than our NMP system in terms of raw speed.
However, a direct comparison between \NMP\ and the supercomputer is not entirely fair.
The objective of \NMP\ is not to surpass the supercomputer's speed but to achieve efficient computation with fewer computing resources.

To compare the computational efficiency of \NMP\ and PaKman on a supercomputer, we examine the throughput under identical resource constraints.
We consider the number of samples assembled in 4,813 seconds using 1,024 {\NMP}s.
Within this timeframe, 1,024 {\NMP}s can process 1,024 human genome assemblies, whereas the supercomputer can process only 123 human assemblies in the same timeframe. 
This demonstrates that \NMP\ can perform \THROUGHPUT\ more assemblies than the supercomputer under the same resource constraints. 

Adopting \NMP\ for use in supercomputers could be another direction for improving genome assembly performance.
Given that Iterative Compaction takes 63\% of the total runtime on the supercomputer \cite{ghosh2020pakman}, integrating \NMP\ into the supercomputer framework could result in a 2.46$\times$ performance gain.










\subsection{Area and Power Overhead} \label{640_overhead}

\begin{table}[h!]
\centering 
\resizebox{\columnwidth}{!}{%
    \begin{tabular}{|l|c|c|}
    \hline
    \textbf{} & \textbf{Area (mm\(^2\))} & \textbf{Power (mW)} \\ \hline
    MacroNode Buffer (4 KB) $\times$2 & 0.038 & 9.2 \\ \hline
    TransferNode Scratchpad (1 KB) $\times$2 & 0.009 & 2.3 \\ \hline
    ALU $\times$3 & 0.037 & 18.5 \\ \hline
    Crossbar Switch & 0.025 & 0.3 \\ \hline
    \textbf{PE} & \textbf{0.110} & \textbf{30.6} \\ \hline
    \textbf{16 PEs} & \textbf{1.763} & \textbf{489.3} \\ \hline
    \end{tabular}
}
\caption{Area Overhead and Power Consumption}
\vspace{-7mm}
\label{tab:area_power}
\end{table}

Using post-synthesis results from a commercial 28 nm technology node, we evaluate \NMP's area overhead and power consumption following methodologies from prior works~\cite{talati2022ndminer,talati2022mint}.
Each PE occupies $0.11 mm^2$ and draws $30.6 mW$ of power, with component-level breakdowns presented in Table \ref{tab:area_power}.
Compared to the typical buffer chip area of $100mm^2$ and a single DIMM's power consumption of $13W$ \cite{meaney2015ibm, ke2020recnmp,kwon2019tensordimm}, \NMP\ with 16 PEs incurs negligible overhead: 1.8\% in area and 3.8\% in power.

\subsection{Comparison with GPUs} \label{650_gpu}


\added{GPUs are promising for many applications due to their superior performance, driven by massive parallelism and high memory bandwidth. However, they are constrained by limited on-device memory capacity.}
The limited memory capacity (80 GB) of state-of-the-art GPUs, such as the H100 and A100, makes them unsuitable for high-quality large-scale genome assembly.
Memory constraints necessitate significant reductions in batch sizes, which severely degrade the quality of generated genome contigs.
Our experiments show that maintaining a memory footprint below 80 GB restricts the batch size to less than 4\% of the human genome, resulting in poor contig quality (N50 = 1200) as shown in Table \ref{tab:n50}.
Achieving acceptable contig quality (N50 = 3000) requires at least 350 GB of memory, far exceeding current GPU memory capacities.

\added{Using a GPU cluster could potentially mitigate the memory capacity limitation. 
However, such a cluster requires substantial computational resources.
Accommodating a 379 GB memory footprint would necessitate five NVIDIA A100 80 GB GPUs, consuming 1500 W of power and occupying 4130 $mm^2$ of die area~\cite{gu2023gendp}.
In contrast, {\NMP} with an 8-DIMM 512 GB configuration consumes 385$\times$ less power and occupies 293$\times$ less die area (3.9 W, 14.1 $mm^2$).
}


The limitations of using GPUs in \textit{de novo} assembly are supported by a recent study \cite{zhou2021ultra} indicating that GPU-based de Bruijn graph assemblers are not recommended due to limited applicability and performance, particularly for large-scale genome assembly.
The Megahit developers \cite{li2015megahit} also report that GPU-based implementations underperform compared to CPU versions while introducing significant usability challenges \cite{zhou2021ultra}.

\section{Related Works} \label{800_related_work}



\paragraph{\textbf{In/Near-memory processing accelerators for \textit{de novo} assembly}}

Previous work has explored accelerating De Bruijn graph-based genome assembly using in/near-memory processing. 
Zhou et al. \cite{zhou2021ultra} enhanced the Megahit assembly algorithm \cite{li2015megahit} through the use of hybrid memory cubes, exploiting high degrees of parallelism and enhanced memory bandwidth across near-data processing cores. 
However, the Megahit algorithm does not scale well with increasing genome sample data sizes, and hybrid memory cubes have been discontinued since 2018.

Angizi et al. \cite{angizi2020pim} introduced PIM-Assembler, a processing-in-DRAM platform for \textit{de novo} assembly.
Their design achieves high performance and energy efficiency through in-DRAM X(N)OR operations for accelerating comparisons and additions, along with optimized data partitioning and mapping techniques.
However, while effective for smaller genomes (9.2 GB), their approach has not been demonstrated for extremely large-scale \textit{de novo} assembly tasks that we address in our work.

While our work focuses on accelerating the Iterative Compaction step, Wu et al. \cite{wu2024abakus} accelerated the k-mer counting step using in-storage processing. 
Their method leverages the embedded CPU controllers and DRAM within the solid-state drive, thereby reducing data transfer between the SSD and the main memory.
In contrast, \NMP\ aims to speed up the Iterative Compaction step, which is the most time-consuming phase.

\paragraph{\textbf{GPU-based \textit{de novo} assembly accelerators}}

Several works \cite{awan2021accelerating,goswami2018gpu,mahmood2011gpu,jain2013gagm,swiercz2018grasshopper} leveraged GPUs' parallel processing capabilities and high bandwidth to accelerate specific assembly operations, such as local assembly, sorting, and prefix scan.

However, GPUs face significant limitations for large-scale genome assembly. 
Their restricted onboard memory capacity \cite{zhou2021ultra} makes them inadequate for extremely large datasets. 
Additionally, genome assembly algorithms present challenges for GPUs due to their memory-intensive nature and irregular access patterns during graph construction and traversal \cite{awan2021accelerating}.


\paragraph{\textbf{Non-de Bruijn graph assembly accelerators}}

There are other algorithms that researchers use when new genomes are discovered.
These algorithms uncover new genomes through sequence mapping and alignment with reference genomes.
BWA-MEM \cite{bwa-mem2} and \texttt{minimap2} \cite{minimap2} are widely used solutions for short-read and long-read alignment, respectively. 


Multiple studies have explored accelerating these algorithms.
\texttt{mm2-fast} \cite{mm2-fast} accelerated the \texttt{minimap2} algorithm on CPUs using SIMD optimizations, while \texttt{mm2-ax} \cite{mm2-ax} took a cooperative approach by designing a heterogeneous system involving both the CPU and GPU. 
\texttt{mm2-gb} \cite{mm2-gb} further improved \texttt{mm2-ax} for long-read sequence mapping by running all the steps on the GPU.
Additionally, Guo et al. \cite{guo2019hardware} proposed a custom accelerator for sequence mapping using FPGA.
Beyond sequence-to-sequence mapping algorithms like \texttt{minimap2}, several accelerators targeted sequence-to-graph mapping of long reads, including SeGraM \cite{cali2022segram} and Harp \cite{zhang2024harp}.

Unlike the above works, \NMP\ addresses genome assembly without the need for a reference genome. 
In addition to performance gains, \NMP\ focuses on reducing the memory footprint and resource requirements for large datasets.

\paragraph{\textbf{Accelerators for other genomic applications}}

MegIS \cite{ghiasi2024megis} addressed the significant data movement overhead in metagenomics analysis through in-storage processing. 
While GenDP \cite{gu2023gendp} and QUETZAL \cite{pavon2024quetzal} accelerated dynamic programming algorithms (Smith-Waterman, Needleman-Wunsch) that are essential kernels in various genomic applications, including reference-guided assembly, \textit{de novo} assembly, metagenomics, and sequence analysis for both long and short reads, our work addresses similar fundamental challenges of processing large, interdependent, and dynamic data structures.
However, we focus on end-to-end \textit{de novo} assembly performance, prioritizing efficient computation with minimal hardware resources while handling large memory requirements.

\section{Conclusion} \label{900_conclusion}

This paper introduced {\NMP}, a novel software-hardware co-optimization framework for accelerating \textit{de novo} genome assembly through near-memory processing.
{\NMP} addresses three critical challenges: a large memory footprint, a memory-latency bottleneck, and irregular data patterns by leveraging near-data processing with customized processing elements, batch processing, and hybrid CPU-NMP processing.
Our evaluation showed that {\NMP} achieves a \FOOTPRINTGAIN\ reduction in memory footprint compared to the state-of-the-art \textit{de novo} assembly.
{\NMP} outperformed the CPU baseline by \NMPGAIN\ and achieved \THROUGHPUT\ greater computational throughput than state-of-the-art \textit{de novo} assembly under the same resource constraints.


\bibliographystyle{ACM-Reference-Format}
\bibliography{refs}

\begin{thebibliography}{58}


\ifx \showCODEN    \undefined \def \showCODEN     #1{\unskip}     \fi
\ifx \showDOI      \undefined \def \showDOI       #1{#1}\fi
\ifx \showISBNx    \undefined \def \showISBNx     #1{\unskip}     \fi
\ifx \showISBNxiii \undefined \def \showISBNxiii  #1{\unskip}     \fi
\ifx \showISSN     \undefined \def \showISSN      #1{\unskip}     \fi
\ifx \showLCCN     \undefined \def \showLCCN      #1{\unskip}     \fi
\ifx \shownote     \undefined \def \shownote      #1{#1}          \fi
\ifx \showarticletitle \undefined \def \showarticletitle #1{#1}   \fi
\ifx \showURL      \undefined \def \showURL       {\relax}        \fi
\providecommand\bibfield[2]{#2}
\providecommand\bibinfo[2]{#2}
\providecommand\natexlab[1]{#1}
\providecommand\showeprint[2][]{arXiv:#2}

\bibitem[hum(2009)]%
        {human_genome}
 \bibinfo{year}{2009}\natexlab{}.
\newblock \bibinfo{title}{{Homo sapiens genome}}.
\newblock \bibinfo{howpublished}{\url{https://www.ncbi.nlm.nih.gov/datasets/genome/GCF_000001405.13/}}.
\newblock
\newblock
\shownote{[Online; accessed 18-Nov-2024]}.


\bibitem[ssd(2020)]%
        {ssd_bw}
 \bibinfo{year}{2020}\natexlab{}.
\newblock \bibinfo{title}{{980 PRO}}.
\newblock \bibinfo{howpublished}{\url{https://semiconductor.samsung.com/consumer-storage/internal-ssd/980pro/}}.
\newblock
\newblock
\shownote{[Online; accessed 10-Feb-2025]}.


\bibitem[Angizi et~al\mbox{.}(2020)]%
        {angizi2020pim}
\bibfield{author}{\bibinfo{person}{Shaahin Angizi}, \bibinfo{person}{Naima~Ahmed Fahmi}, \bibinfo{person}{Wei Zhang}, {and} \bibinfo{person}{Deliang Fan}.} \bibinfo{year}{2020}\natexlab{}.
\newblock \showarticletitle{Pim-assembler: A processing-in-memory platform for genome assembly}. In \bibinfo{booktitle}{\emph{2020 57th ACM/IEEE design automation conference (DAC)}}. IEEE, \bibinfo{pages}{1--6}.
\newblock


\bibitem[Awan et~al\mbox{.}(2021)]%
        {awan2021accelerating}
\bibfield{author}{\bibinfo{person}{Muaaz~Gul Awan}, \bibinfo{person}{Steven Hofmeyr}, \bibinfo{person}{Rob Egan}, \bibinfo{person}{Nan Ding}, \bibinfo{person}{Aydin Buluc}, \bibinfo{person}{Jack Deslippe}, \bibinfo{person}{Leonid Oliker}, {and} \bibinfo{person}{Katherine Yelick}.} \bibinfo{year}{2021}\natexlab{}.
\newblock \showarticletitle{Accelerating large scale de novo metagenome assembly using GPUs}. In \bibinfo{booktitle}{\emph{Proceedings of the International Conference for High Performance Computing, Networking, Storage and Analysis}}. \bibinfo{pages}{1--11}.
\newblock


\bibitem[Butler et~al\mbox{.}(2008)]%
        {butler2008allpaths}
\bibfield{author}{\bibinfo{person}{Jonathan Butler}, \bibinfo{person}{Iain MacCallum}, \bibinfo{person}{Michael Kleber}, \bibinfo{person}{Ilya~A Shlyakhter}, \bibinfo{person}{Matthew~K Belmonte}, \bibinfo{person}{Eric~S Lander}, \bibinfo{person}{Chad Nusbaum}, {and} \bibinfo{person}{David~B Jaffe}.} \bibinfo{year}{2008}\natexlab{}.
\newblock \showarticletitle{ALLPATHS: de novo assembly of whole-genome shotgun microreads}.
\newblock \bibinfo{journal}{\emph{Genome research}} \bibinfo{volume}{18}, \bibinfo{number}{5} (\bibinfo{year}{2008}), \bibinfo{pages}{810--820}.
\newblock


\bibitem[Cali et~al\mbox{.}(2022)]%
        {cali2022segram}
\bibfield{author}{\bibinfo{person}{Damla~Senol Cali}, \bibinfo{person}{Konstantinos Kanellopoulos}, \bibinfo{person}{Jo{\"e}l Lindegger}, \bibinfo{person}{Z{\"u}lal Bing{\"o}l}, \bibinfo{person}{Gurpreet~S Kalsi}, \bibinfo{person}{Ziyi Zuo}, \bibinfo{person}{Can Firtina}, \bibinfo{person}{Meryem~Banu Cavlak}, \bibinfo{person}{Jeremie Kim}, \bibinfo{person}{Nika~Mansouri Ghiasi}, {et~al\mbox{.}}} \bibinfo{year}{2022}\natexlab{}.
\newblock \showarticletitle{SeGraM: A universal hardware accelerator for genomic sequence-to-graph and sequence-to-sequence mapping}. In \bibinfo{booktitle}{\emph{Proceedings of the 49th Annual International Symposium on Computer Architecture}}. \bibinfo{pages}{638--655}.
\newblock


\bibitem[Carlson et~al\mbox{.}(2014)]%
        {carlson2014aeohmcm}
\bibfield{author}{\bibinfo{person}{Trevor~E. Carlson}, \bibinfo{person}{Wim Heirman}, \bibinfo{person}{Stijn Eyerman}, \bibinfo{person}{Ibrahim Hur}, {and} \bibinfo{person}{Lieven Eeckhout}.} \bibinfo{year}{2014}\natexlab{}.
\newblock \showarticletitle{An Evaluation of High-Level Mechanistic Core Models}.
\newblock \bibinfo{journal}{\emph{ACM Transactions on Architecture and Code Optimization (TACO)}}, Article \bibinfo{articleno}{5} (\bibinfo{year}{2014}), \bibinfo{numpages}{23}~pages.
\newblock
\showISSN{1544-3566}
\urldef\tempurl%
\url{https://doi.org/10.1145/2629677}
\showDOI{\tempurl}


\bibitem[Cheng et~al\mbox{.}(2021)]%
        {cheng2021haplotype}
\bibfield{author}{\bibinfo{person}{Haoyu Cheng}, \bibinfo{person}{Gregory~T Concepcion}, \bibinfo{person}{Xiaowen Feng}, \bibinfo{person}{Haowen Zhang}, {and} \bibinfo{person}{Heng Li}.} \bibinfo{year}{2021}\natexlab{}.
\newblock \showarticletitle{Haplotype-resolved de novo assembly using phased assembly graphs with hifiasm}.
\newblock \bibinfo{journal}{\emph{Nature methods}} \bibinfo{volume}{18}, \bibinfo{number}{2} (\bibinfo{year}{2021}), \bibinfo{pages}{170--175}.
\newblock


\bibitem[Devaux(2019)]%
        {devaux2019true}
\bibfield{author}{\bibinfo{person}{Fabrice Devaux}.} \bibinfo{year}{2019}\natexlab{}.
\newblock \showarticletitle{The true processing in memory accelerator}. In \bibinfo{booktitle}{\emph{2019 IEEE Hot Chips 31 Symposium (HCS)}}. IEEE Computer Society, \bibinfo{pages}{1--24}.
\newblock


\bibitem[Dong et~al\mbox{.}(2024)]%
        {mm2-gb}
\bibfield{author}{\bibinfo{person}{Juechu Dong}, \bibinfo{person}{Xueshen Liu}, \bibinfo{person}{Harisankar Sadasivan}, \bibinfo{person}{Sriranjani Sitaraman}, {and} \bibinfo{person}{Satish Narayanasamy}.} \bibinfo{year}{2024}\natexlab{}.
\newblock \showarticletitle{mm2-gb: GPU Accelerated Minimap2 for Long Read DNA Mapping}.
\newblock \bibinfo{journal}{\emph{bioRxiv}} (\bibinfo{year}{2024}), \bibinfo{pages}{2024--03}.
\newblock


\bibitem[Donia et~al\mbox{.}(2014)]%
        {donia2014systematic}
\bibfield{author}{\bibinfo{person}{Mohamed~S Donia}, \bibinfo{person}{Peter Cimermancic}, \bibinfo{person}{Christopher~J Schulze}, \bibinfo{person}{Laura C~Wieland Brown}, \bibinfo{person}{John Martin}, \bibinfo{person}{Makedonka Mitreva}, \bibinfo{person}{Jon Clardy}, \bibinfo{person}{Roger~G Linington}, {and} \bibinfo{person}{Michael~A Fischbach}.} \bibinfo{year}{2014}\natexlab{}.
\newblock \showarticletitle{A systematic analysis of biosynthetic gene clusters in the human microbiome reveals a common family of antibiotics}.
\newblock \bibinfo{journal}{\emph{Cell}} \bibinfo{volume}{158}, \bibinfo{number}{6} (\bibinfo{year}{2014}), \bibinfo{pages}{1402--1414}.
\newblock


\bibitem[Ekim et~al\mbox{.}(2021)]%
        {ekim2021minimizer}
\bibfield{author}{\bibinfo{person}{Bar{\i}{\c{s}} Ekim}, \bibinfo{person}{Bonnie Berger}, {and} \bibinfo{person}{Rayan Chikhi}.} \bibinfo{year}{2021}\natexlab{}.
\newblock \showarticletitle{Minimizer-space de Bruijn graphs: Whole-genome assembly of long reads in minutes on a personal computer}.
\newblock \bibinfo{journal}{\emph{Cell systems}} \bibinfo{volume}{12}, \bibinfo{number}{10} (\bibinfo{year}{2021}), \bibinfo{pages}{958--968}.
\newblock


\bibitem[Feng et~al\mbox{.}(2022)]%
        {feng2022menda}
\bibfield{author}{\bibinfo{person}{Siying Feng}, \bibinfo{person}{Xin He}, \bibinfo{person}{Kuan-Yu Chen}, \bibinfo{person}{Liu Ke}, \bibinfo{person}{Xuan Zhang}, \bibinfo{person}{David Blaauw}, \bibinfo{person}{Trevor Mudge}, {and} \bibinfo{person}{Ronald Dreslinski}.} \bibinfo{year}{2022}\natexlab{}.
\newblock \showarticletitle{MeNDA: a near-memory multi-way merge solution for sparse transposition and dataflows}. In \bibinfo{booktitle}{\emph{Proceedings of the 49th Annual International Symposium on Computer Architecture}}. \bibinfo{pages}{245--258}.
\newblock


\bibitem[Fujiki et~al\mbox{.}(2021)]%
        {fujiki2021near}
\bibfield{author}{\bibinfo{person}{Daichi Fujiki}, \bibinfo{person}{Xiaowei Wang}, \bibinfo{person}{Arun Subramaniyan}, {and} \bibinfo{person}{Reetuparna Das}.} \bibinfo{year}{2021}\natexlab{}.
\newblock \showarticletitle{In-/near-memory Computing}.
\newblock \bibinfo{journal}{\emph{Synthesis Lectures on Computer Architecture}} \bibinfo{volume}{16}, \bibinfo{number}{2} (\bibinfo{year}{2021}), \bibinfo{pages}{1--140}.
\newblock


\bibitem[Georganas et~al\mbox{.}(2015)]%
        {georganas2015hipmer}
\bibfield{author}{\bibinfo{person}{Evangelos Georganas}, \bibinfo{person}{Ayd{\i}n Bulu{\c{c}}}, \bibinfo{person}{Jarrod Chapman}, \bibinfo{person}{Steven Hofmeyr}, \bibinfo{person}{Chaitanya Aluru}, \bibinfo{person}{Rob Egan}, \bibinfo{person}{Leonid Oliker}, \bibinfo{person}{Daniel Rokhsar}, {and} \bibinfo{person}{Katherine Yelick}.} \bibinfo{year}{2015}\natexlab{}.
\newblock \showarticletitle{HipMer: an extreme-scale de novo genome assembler}. In \bibinfo{booktitle}{\emph{Proceedings of the International Conference for High Performance Computing, Networking, Storage and Analysis}}. \bibinfo{pages}{1--11}.
\newblock


\bibitem[Georganas et~al\mbox{.}(2018)]%
        {georganas2018extreme}
\bibfield{author}{\bibinfo{person}{Evangelos Georganas}, \bibinfo{person}{Rob Egan}, \bibinfo{person}{Steven Hofmeyr}, \bibinfo{person}{Eugene Goltsman}, \bibinfo{person}{Bill Arndt}, \bibinfo{person}{Andrew Tritt}, \bibinfo{person}{Aydin Bulu{\c{c}}}, \bibinfo{person}{Leonid Oliker}, {and} \bibinfo{person}{Katherine Yelick}.} \bibinfo{year}{2018}\natexlab{}.
\newblock \showarticletitle{Extreme scale de novo metagenome assembly}. In \bibinfo{booktitle}{\emph{SC18: International Conference for High Performance Computing, Networking, Storage and Analysis}}. IEEE, \bibinfo{pages}{122--134}.
\newblock


\bibitem[Ghiasi et~al\mbox{.}(2024)]%
        {ghiasi2024megis}
\bibfield{author}{\bibinfo{person}{Nika~Mansouri Ghiasi}, \bibinfo{person}{Mohammad Sadrosadati}, \bibinfo{person}{Harun Mustafa}, \bibinfo{person}{Arvid Gollwitzer}, \bibinfo{person}{Can Firtina}, \bibinfo{person}{Julien Eudine}, \bibinfo{person}{Haiyu Mao}, \bibinfo{person}{Jo{\"e}l Lindegger}, \bibinfo{person}{Meryem~Banu Cavlak}, \bibinfo{person}{Mohammed Alser}, {et~al\mbox{.}}} \bibinfo{year}{2024}\natexlab{}.
\newblock \showarticletitle{MegIS: High-Performance, Energy-Efficient, and Low-Cost Metagenomic Analysis with In-Storage Processing}. In \bibinfo{booktitle}{\emph{2024 ACM/IEEE 51st Annual International Symposium on Computer Architecture (ISCA)}}. IEEE, \bibinfo{pages}{660--677}.
\newblock


\bibitem[Ghosh et~al\mbox{.}(2020)]%
        {ghosh2020pakman}
\bibfield{author}{\bibinfo{person}{Priyanka Ghosh}, \bibinfo{person}{Sriram Krishnamoorthy}, {and} \bibinfo{person}{Ananth Kalyanaraman}.} \bibinfo{year}{2020}\natexlab{}.
\newblock \showarticletitle{Pakman: a scalable algorithm for generating genomic contigs on distributed memory machines}.
\newblock \bibinfo{journal}{\emph{IEEE Transactions on Parallel and Distributed Systems}} \bibinfo{volume}{32}, \bibinfo{number}{5} (\bibinfo{year}{2020}), \bibinfo{pages}{1191--1209}.
\newblock


\bibitem[G{\'o}mez-Luna et~al\mbox{.}(2022)]%
        {gomez2022benchmarking}
\bibfield{author}{\bibinfo{person}{Juan G{\'o}mez-Luna}, \bibinfo{person}{Izzat El~Hajj}, \bibinfo{person}{Ivan Fernandez}, \bibinfo{person}{Christina Giannoula}, \bibinfo{person}{Geraldo~F Oliveira}, {and} \bibinfo{person}{Onur Mutlu}.} \bibinfo{year}{2022}\natexlab{}.
\newblock \showarticletitle{Benchmarking a new paradigm: Experimental analysis and characterization of a real processing-in-memory system}.
\newblock \bibinfo{journal}{\emph{IEEE Access}}  \bibinfo{volume}{10} (\bibinfo{year}{2022}), \bibinfo{pages}{52565--52608}.
\newblock


\bibitem[Goswami et~al\mbox{.}(2018)]%
        {goswami2018gpu}
\bibfield{author}{\bibinfo{person}{Sayan Goswami}, \bibinfo{person}{Kisung Lee}, \bibinfo{person}{Shayan Shams}, {and} \bibinfo{person}{Seung-Jong Park}.} \bibinfo{year}{2018}\natexlab{}.
\newblock \showarticletitle{Gpu-accelerated large-scale genome assembly}. In \bibinfo{booktitle}{\emph{2018 IEEE International Parallel and Distributed Processing Symposium (IPDPS)}}. IEEE, \bibinfo{pages}{814--824}.
\newblock


\bibitem[Gu et~al\mbox{.}(2023)]%
        {gu2023gendp}
\bibfield{author}{\bibinfo{person}{Yufeng Gu}, \bibinfo{person}{Arun Subramaniyan}, \bibinfo{person}{Tim Dunn}, \bibinfo{person}{Alireza Khadem}, \bibinfo{person}{Kuan-Yu Chen}, \bibinfo{person}{Somnath Paul}, \bibinfo{person}{Md Vasimuddin}, \bibinfo{person}{Sanchit Misra}, \bibinfo{person}{David Blaauw}, \bibinfo{person}{Satish Narayanasamy}, {et~al\mbox{.}}} \bibinfo{year}{2023}\natexlab{}.
\newblock \showarticletitle{GenDP: A Framework of Dynamic Programming Acceleration for Genome Sequencing Analysis}. In \bibinfo{booktitle}{\emph{Proceedings of the 50th Annual International Symposium on Computer Architecture}}. \bibinfo{pages}{1--15}.
\newblock


\bibitem[Guo et~al\mbox{.}(2019)]%
        {guo2019hardware}
\bibfield{author}{\bibinfo{person}{Licheng Guo}, \bibinfo{person}{Jason Lau}, \bibinfo{person}{Zhenyuan Ruan}, \bibinfo{person}{Peng Wei}, {and} \bibinfo{person}{Jason Cong}.} \bibinfo{year}{2019}\natexlab{}.
\newblock \showarticletitle{Hardware acceleration of long read pairwise overlapping in genome sequencing: A race between fpga and gpu}. In \bibinfo{booktitle}{\emph{2019 IEEE 27th Annual International Symposium on Field-Programmable Custom Computing Machines (FCCM)}}. IEEE, \bibinfo{pages}{127--135}.
\newblock


\bibitem[Gurevich et~al\mbox{.}(2013)]%
        {gurevich2013quast}
\bibfield{author}{\bibinfo{person}{Alexey Gurevich}, \bibinfo{person}{Vladislav Saveliev}, \bibinfo{person}{Nikolay Vyahhi}, {and} \bibinfo{person}{Glenn Tesler}.} \bibinfo{year}{2013}\natexlab{}.
\newblock \showarticletitle{QUAST: quality assessment tool for genome assemblies}.
\newblock \bibinfo{journal}{\emph{Bioinformatics}} \bibinfo{volume}{29}, \bibinfo{number}{8} (\bibinfo{year}{2013}), \bibinfo{pages}{1072--1075}.
\newblock


\bibitem[Huang et~al\mbox{.}(2012)]%
        {huang2012art}
\bibfield{author}{\bibinfo{person}{Weichun Huang}, \bibinfo{person}{Leping Li}, \bibinfo{person}{Jason~R Myers}, {and} \bibinfo{person}{Gabor~T Marth}.} \bibinfo{year}{2012}\natexlab{}.
\newblock \showarticletitle{ART: a next-generation sequencing read simulator}.
\newblock \bibinfo{journal}{\emph{Bioinformatics}} \bibinfo{volume}{28}, \bibinfo{number}{4} (\bibinfo{year}{2012}), \bibinfo{pages}{593--594}.
\newblock


\bibitem[Jain et~al\mbox{.}(2013)]%
        {jain2013gagm}
\bibfield{author}{\bibinfo{person}{Ashutosh Jain}, \bibinfo{person}{Anshuj Garg}, {and} \bibinfo{person}{Kolin Paul}.} \bibinfo{year}{2013}\natexlab{}.
\newblock \showarticletitle{GAGM: Genome assembly on GPU using mate pairs}. In \bibinfo{booktitle}{\emph{20th Annual International Conference on High Performance Computing}}. IEEE, \bibinfo{pages}{176--185}.
\newblock


\bibitem[Kalikar et~al\mbox{.}(2022)]%
        {mm2-fast}
\bibfield{author}{\bibinfo{person}{Saurabh Kalikar}, \bibinfo{person}{Chirag Jain}, \bibinfo{person}{Md Vasimuddin}, {and} \bibinfo{person}{Sanchit Misra}.} \bibinfo{year}{2022}\natexlab{}.
\newblock \showarticletitle{Accelerating minimap2 for long-read sequencing applications on modern CPUs}.
\newblock \bibinfo{journal}{\emph{Nature Computational Science}} \bibinfo{volume}{2}, \bibinfo{number}{2} (\bibinfo{year}{2022}), \bibinfo{pages}{78--83}.
\newblock


\bibitem[Ke et~al\mbox{.}(2020)]%
        {ke2020recnmp}
\bibfield{author}{\bibinfo{person}{Liu Ke}, \bibinfo{person}{Udit Gupta}, \bibinfo{person}{Benjamin~Youngjae Cho}, \bibinfo{person}{David Brooks}, \bibinfo{person}{Vikas Chandra}, \bibinfo{person}{Utku Diril}, \bibinfo{person}{Amin Firoozshahian}, \bibinfo{person}{Kim Hazelwood}, \bibinfo{person}{Bill Jia}, \bibinfo{person}{Hsien-Hsin~S Lee}, \bibinfo{person}{Meng Li}, \bibinfo{person}{Bert Maher}, \bibinfo{person}{Dheevatsa Mudigere}, \bibinfo{person}{Maxim Naumov}, \bibinfo{person}{Martin Schatz}, \bibinfo{person}{Mikhail Smelyanskiy}, \bibinfo{person}{Xiaodong Wang}, \bibinfo{person}{Brandon Reagen}, \bibinfo{person}{Carole-Jean Wu}, \bibinfo{person}{Mark Hempstead}, {and} \bibinfo{person}{Xuan Zhang}.} \bibinfo{year}{2020}\natexlab{}.
\newblock \showarticletitle{Recnmp: Accelerating personalized recommendation with near-memory processing}. In \bibinfo{booktitle}{\emph{ISCA}}. \bibinfo{pages}{790--803}.
\newblock


\bibitem[Kim et~al\mbox{.}(2023)]%
        {kim2023recpim}
\bibfield{author}{\bibinfo{person}{Heewoo Kim}, \bibinfo{person}{Haojie Ye}, \bibinfo{person}{Trevor Mudge}, \bibinfo{person}{Ronald Dreslinski}, {and} \bibinfo{person}{Nishil Talati}.} \bibinfo{year}{2023}\natexlab{}.
\newblock \showarticletitle{RecPIM: A PIM-Enabled DRAM-RRAM Hybrid Memory System For Recommendation Models}. In \bibinfo{booktitle}{\emph{2023 IEEE/ACM International Symposium on Low Power Electronics and Design (ISLPED)}}. IEEE, \bibinfo{pages}{1--6}.
\newblock


\bibitem[Kim et~al\mbox{.}(2021)]%
        {kim2021aquabolt}
\bibfield{author}{\bibinfo{person}{Jin~Hyun Kim}, \bibinfo{person}{Shin-haeng Kang}, \bibinfo{person}{Sukhan Lee}, \bibinfo{person}{Hyeonsu Kim}, \bibinfo{person}{Woongjae Song}, \bibinfo{person}{Yuhwan Ro}, \bibinfo{person}{Seungwon Lee}, \bibinfo{person}{David Wang}, \bibinfo{person}{Hyunsung Shin}, \bibinfo{person}{Bengseng Phuah}, \bibinfo{person}{Jihyun Choi}, \bibinfo{person}{Jinin So}, \bibinfo{person}{YeonGon Cho}, \bibinfo{person}{JoonHo Song}, \bibinfo{person}{Jangseok Choi}, \bibinfo{person}{Jeonghyeon Cho}, \bibinfo{person}{Kyomin Sohn}, \bibinfo{person}{Youngsoo Sohn}, \bibinfo{person}{Kwangil Park}, {and} \bibinfo{person}{Nam~Sung Kim}.} \bibinfo{year}{2021}\natexlab{}.
\newblock \showarticletitle{Aquabolt-XL: Samsung HBM2-PIM with in-memory processing for ML accelerators and beyond}. In \bibinfo{booktitle}{\emph{2021 IEEE Hot Chips 33 Symposium (HCS)}}. IEEE, \bibinfo{pages}{1--26}.
\newblock


\bibitem[Kim et~al\mbox{.}(2015)]%
        {kim2015ramulator}
\bibfield{author}{\bibinfo{person}{Yoongu Kim}, \bibinfo{person}{Weikun Yang}, {and} \bibinfo{person}{Onur Mutlu}.} \bibinfo{year}{2015}\natexlab{}.
\newblock \showarticletitle{Ramulator: A fast and extensible DRAM simulator}.
\newblock \bibinfo{journal}{\emph{IEEE Computer architecture letters}} \bibinfo{volume}{15}, \bibinfo{number}{1} (\bibinfo{year}{2015}), \bibinfo{pages}{45--49}.
\newblock


\bibitem[Koren et~al\mbox{.}(2017)]%
        {koren2017canu}
\bibfield{author}{\bibinfo{person}{Sergey Koren}, \bibinfo{person}{Brian~P Walenz}, \bibinfo{person}{Konstantin Berlin}, \bibinfo{person}{Jason~R Miller}, \bibinfo{person}{Nicholas~H Bergman}, {and} \bibinfo{person}{Adam~M Phillippy}.} \bibinfo{year}{2017}\natexlab{}.
\newblock \showarticletitle{Canu: scalable and accurate long-read assembly via adaptive k-mer weighting and repeat separation}.
\newblock \bibinfo{journal}{\emph{Genome research}} \bibinfo{volume}{27}, \bibinfo{number}{5} (\bibinfo{year}{2017}), \bibinfo{pages}{722--736}.
\newblock


\bibitem[Kwon et~al\mbox{.}(2019)]%
        {kwon2019tensordimm}
\bibfield{author}{\bibinfo{person}{Youngeun Kwon}, \bibinfo{person}{Yunjae Lee}, {and} \bibinfo{person}{Minsoo Rhu}.} \bibinfo{year}{2019}\natexlab{}.
\newblock \showarticletitle{Tensordimm: A practical near-memory processing architecture for embeddings and tensor operations in deep learning}. In \bibinfo{booktitle}{\emph{Proceedings of the 52nd Annual IEEE/ACM International Symposium on Microarchitecture}}. \bibinfo{pages}{740--753}.
\newblock


\bibitem[Lee et~al\mbox{.}(2021)]%
        {lee2021hardware}
\bibfield{author}{\bibinfo{person}{Sukhan Lee}, \bibinfo{person}{Shin-haeng Kang}, \bibinfo{person}{Jaehoon Lee}, \bibinfo{person}{Hyeonsu Kim}, \bibinfo{person}{Eojin Lee}, \bibinfo{person}{Seungwoo Seo}, \bibinfo{person}{Hosang Yoon}, \bibinfo{person}{Seungwon Lee}, \bibinfo{person}{Kyounghwan Lim}, \bibinfo{person}{Hyunsung Shin}, \bibinfo{person}{Jinhyun Kim}, \bibinfo{person}{Seongil O}, \bibinfo{person}{Anand Iyer}, \bibinfo{person}{David Wang}, \bibinfo{person}{Kyomin Sohn}, {and} \bibinfo{person}{Nam~Sung Kim}.} \bibinfo{year}{2021}\natexlab{}.
\newblock \showarticletitle{Hardware architecture and software stack for PIM based on commercial DRAM technology: Industrial product}. In \bibinfo{booktitle}{\emph{2021 ACM/IEEE 48th Annual International Symposium on Computer Architecture (ISCA)}}. IEEE, \bibinfo{pages}{43--56}.
\newblock


\bibitem[Li et~al\mbox{.}(2015)]%
        {li2015megahit}
\bibfield{author}{\bibinfo{person}{Dinghua Li}, \bibinfo{person}{Chi-Man Liu}, \bibinfo{person}{Ruibang Luo}, \bibinfo{person}{Kunihiko Sadakane}, {and} \bibinfo{person}{Tak-Wah Lam}.} \bibinfo{year}{2015}\natexlab{}.
\newblock \showarticletitle{MEGAHIT: an ultra-fast single-node solution for large and complex metagenomics assembly via succinct de Bruijn graph}.
\newblock \bibinfo{journal}{\emph{Bioinformatics}} \bibinfo{volume}{31}, \bibinfo{number}{10} (\bibinfo{year}{2015}), \bibinfo{pages}{1674--1676}.
\newblock


\bibitem[Li(2018)]%
        {minimap2}
\bibfield{author}{\bibinfo{person}{Heng Li}.} \bibinfo{year}{2018}\natexlab{}.
\newblock \showarticletitle{{Minimap2: pairwise alignment for nucleotide sequences}}.
\newblock \bibinfo{journal}{\emph{Bioinformatics}} \bibinfo{volume}{34}, \bibinfo{number}{18} (\bibinfo{date}{05} \bibinfo{year}{2018}), \bibinfo{pages}{3094--3100}.
\newblock
\showISSN{1367-4803}
\urldef\tempurl%
\url{https://doi.org/10.1093/bioinformatics/bty191}
\showDOI{\tempurl}
\showeprint{https://academic.oup.com/bioinformatics/article-pdf/34/18/3094/48919122/bioinformatics\_34\_18\_3094.pdf}


\bibitem[Li et~al\mbox{.}(2016)]%
        {li2016pinatubo}
\bibfield{author}{\bibinfo{person}{Shuangchen Li}, \bibinfo{person}{Cong Xu}, \bibinfo{person}{Qiaosha Zou}, \bibinfo{person}{Jishen Zhao}, \bibinfo{person}{Yu Lu}, {and} \bibinfo{person}{Yuan Xie}.} \bibinfo{year}{2016}\natexlab{}.
\newblock \showarticletitle{Pinatubo: A processing-in-memory architecture for bulk bitwise operations in emerging non-volatile memories}. In \bibinfo{booktitle}{\emph{DAC}}. \bibinfo{pages}{1--6}.
\newblock


\bibitem[Mahmood and Rangwala(2011)]%
        {mahmood2011gpu}
\bibfield{author}{\bibinfo{person}{Syed~Faraz Mahmood} {and} \bibinfo{person}{Huzefa Rangwala}.} \bibinfo{year}{2011}\natexlab{}.
\newblock \showarticletitle{Gpu-euler: Sequence assembly using gpgpu}. In \bibinfo{booktitle}{\emph{2011 IEEE International Conference on High Performance Computing and Communications}}. IEEE, \bibinfo{pages}{153--160}.
\newblock


\bibitem[Meaney et~al\mbox{.}(2015)]%
        {meaney2015ibm}
\bibfield{author}{\bibinfo{person}{Patrick~J Meaney}, \bibinfo{person}{Lawrence~D Curley}, \bibinfo{person}{Glenn~D Gilda}, \bibinfo{person}{Mark~R Hodges}, \bibinfo{person}{Daniel~J Buerkle}, \bibinfo{person}{Robert~D Siegl}, {and} \bibinfo{person}{Roger~K Dong}.} \bibinfo{year}{2015}\natexlab{}.
\newblock \showarticletitle{The IBM z13 memory subsystem for big data}.
\newblock \bibinfo{journal}{\emph{IBM Journal of Research and Development}} \bibinfo{volume}{59}, \bibinfo{number}{4/5} (\bibinfo{year}{2015}), \bibinfo{pages}{4--1}.
\newblock


\bibitem[Pavon et~al\mbox{.}(2024)]%
        {pavon2024quetzal}
\bibfield{author}{\bibinfo{person}{Julian Pavon}, \bibinfo{person}{Ivan~Vargas Valdivieso}, \bibinfo{person}{Carlos Rojas}, \bibinfo{person}{Cesar Hernandez}, \bibinfo{person}{Mehmet Aslan}, \bibinfo{person}{Roger Figueras}, \bibinfo{person}{Yichao Yuan}, \bibinfo{person}{Jo{\"e}l Lindegger}, \bibinfo{person}{Mohammed Alser}, \bibinfo{person}{Francesc Moll}, {et~al\mbox{.}}} \bibinfo{year}{2024}\natexlab{}.
\newblock \showarticletitle{QUETZAL: Vector Acceleration Framework for Modern Genome Sequence Analysis Algorithms}. In \bibinfo{booktitle}{\emph{2024 ACM/IEEE 51st Annual International Symposium on Computer Architecture (ISCA)}}. IEEE, \bibinfo{pages}{597--612}.
\newblock


\bibitem[Peng et~al\mbox{.}(2012)]%
        {peng2012idba}
\bibfield{author}{\bibinfo{person}{Yu Peng}, \bibinfo{person}{Henry~CM Leung}, \bibinfo{person}{Siu-Ming Yiu}, {and} \bibinfo{person}{Francis~YL Chin}.} \bibinfo{year}{2012}\natexlab{}.
\newblock \showarticletitle{IDBA-UD: a de novo assembler for single-cell and metagenomic sequencing data with highly uneven depth}.
\newblock \bibinfo{journal}{\emph{Bioinformatics}} \bibinfo{volume}{28}, \bibinfo{number}{11} (\bibinfo{year}{2012}), \bibinfo{pages}{1420--1428}.
\newblock


\bibitem[Pereira(2019)]%
        {pereira2019metagenomics}
\bibfield{author}{\bibinfo{person}{Flory Pereira}.} \bibinfo{year}{2019}\natexlab{}.
\newblock \showarticletitle{Metagenomics: A gateway to drug discovery}.
\newblock In \bibinfo{booktitle}{\emph{Advances in Biological Science Research}}. \bibinfo{publisher}{Elsevier}, \bibinfo{pages}{453--468}.
\newblock


\bibitem[Sadasivan et~al\mbox{.}(2023)]%
        {mm2-ax}
\bibfield{author}{\bibinfo{person}{Harisankar Sadasivan}, \bibinfo{person}{Milos Maric}, \bibinfo{person}{Eric Dawson}, \bibinfo{person}{Vishanth Iyer}, \bibinfo{person}{Johnny Israeli}, {and} \bibinfo{person}{Satish Narayanasamy}.} \bibinfo{year}{2023}\natexlab{}.
\newblock \showarticletitle{Accelerating Minimap2 for accurate long read alignment on GPUs}.
\newblock \bibinfo{journal}{\emph{Journal of biotechnology and biomedicine}} \bibinfo{volume}{6}, \bibinfo{number}{1} (\bibinfo{year}{2023}), \bibinfo{pages}{13}.
\newblock


\bibitem[Sharon and Banfield(2013)]%
        {sharon2013genomes}
\bibfield{author}{\bibinfo{person}{Itai Sharon} {and} \bibinfo{person}{Jillian~F Banfield}.} \bibinfo{year}{2013}\natexlab{}.
\newblock \showarticletitle{Genomes from metagenomics}.
\newblock \bibinfo{journal}{\emph{Science}} \bibinfo{volume}{342}, \bibinfo{number}{6162} (\bibinfo{year}{2013}), \bibinfo{pages}{1057--1058}.
\newblock


\bibitem[Simpson et~al\mbox{.}(2009)]%
        {simpson2009abyss}
\bibfield{author}{\bibinfo{person}{Jared~T Simpson}, \bibinfo{person}{Kim Wong}, \bibinfo{person}{Shaun~D Jackman}, \bibinfo{person}{Jacqueline~E Schein}, \bibinfo{person}{Steven~JM Jones}, {and} \bibinfo{person}{Inan{\c{c}} Birol}.} \bibinfo{year}{2009}\natexlab{}.
\newblock \showarticletitle{ABySS: a parallel assembler for short read sequence data}.
\newblock \bibinfo{journal}{\emph{Genome research}} \bibinfo{volume}{19}, \bibinfo{number}{6} (\bibinfo{year}{2009}), \bibinfo{pages}{1117--1123}.
\newblock


\bibitem[Su and Naffziger(2023)]%
        {su20231}
\bibfield{author}{\bibinfo{person}{Lisa Su} {and} \bibinfo{person}{Sam Naffziger}.} \bibinfo{year}{2023}\natexlab{}.
\newblock \showarticletitle{1.1 Innovation For the Next Decade of Compute Efficiency}. In \bibinfo{booktitle}{\emph{2023 IEEE International Solid-State Circuits Conference (ISSCC)}}. IEEE, \bibinfo{pages}{8--12}.
\newblock


\bibitem[Swiercz et~al\mbox{.}(2018)]%
        {swiercz2018grasshopper}
\bibfield{author}{\bibinfo{person}{Aleksandra Swiercz}, \bibinfo{person}{Wojciech Frohmberg}, \bibinfo{person}{Michal Kierzynka}, \bibinfo{person}{Pawel Wojciechowski}, \bibinfo{person}{Piotr Zurkowski}, \bibinfo{person}{Jan Badura}, \bibinfo{person}{Artur Laskowski}, \bibinfo{person}{Marta Kasprzak}, {and} \bibinfo{person}{Jacek Blazewicz}.} \bibinfo{year}{2018}\natexlab{}.
\newblock \showarticletitle{GRASShopPER—An algorithm for de novo assembly based on GPU alignments}.
\newblock \bibinfo{journal}{\emph{PloS one}} \bibinfo{volume}{13}, \bibinfo{number}{8} (\bibinfo{year}{2018}), \bibinfo{pages}{e0202355}.
\newblock


\bibitem[Talati et~al\mbox{.}(2016)]%
        {talati2016logic}
\bibfield{author}{\bibinfo{person}{Nishil Talati}, \bibinfo{person}{Saransh Gupta}, \bibinfo{person}{Pravin Mane}, {and} \bibinfo{person}{Shahar Kvatinsky}.} \bibinfo{year}{2016}\natexlab{}.
\newblock \showarticletitle{Logic design within memristive memories using memristor-aided lo{GIC} ({MAGIC})}.
\newblock \bibinfo{journal}{\emph{TNANO}} \bibinfo{volume}{15}, \bibinfo{number}{4} (\bibinfo{year}{2016}), \bibinfo{pages}{635--650}.
\newblock


\bibitem[Talati et~al\mbox{.}(2022a)]%
        {talati2022mint}
\bibfield{author}{\bibinfo{person}{Nishil Talati}, \bibinfo{person}{Haojie Ye}, \bibinfo{person}{Sanketh Vedula}, \bibinfo{person}{Kuan-Yu Chen}, \bibinfo{person}{Yuhan Chen}, \bibinfo{person}{Daniel Liu}, \bibinfo{person}{Yichao Yuan}, \bibinfo{person}{David Blaauw}, \bibinfo{person}{Alex Bronstein}, \bibinfo{person}{Trevor Mudge}, {and} \bibinfo{person}{Ronald Dreslinski}.} \bibinfo{year}{2022}\natexlab{a}.
\newblock \showarticletitle{Mint: An accelerator for mining temporal motifs}. In \bibinfo{booktitle}{\emph{2022 55th IEEE/ACM International Symposium on Microarchitecture (MICRO)}}. IEEE, \bibinfo{pages}{1270--1287}.
\newblock


\bibitem[Talati et~al\mbox{.}(2022b)]%
        {talati2022ndminer}
\bibfield{author}{\bibinfo{person}{Nishil Talati}, \bibinfo{person}{Haojie Ye}, \bibinfo{person}{Yichen Yang}, \bibinfo{person}{Leul Belayneh}, \bibinfo{person}{Kuan-Yu Chen}, \bibinfo{person}{David Blaauw}, \bibinfo{person}{Trevor Mudge}, {and} \bibinfo{person}{Ronald Dreslinski}.} \bibinfo{year}{2022}\natexlab{b}.
\newblock \showarticletitle{Ndminer: accelerating graph pattern mining using near data processing}. In \bibinfo{booktitle}{\emph{Proceedings of the 49th Annual International Symposium on Computer Architecture}}. \bibinfo{pages}{146--159}.
\newblock


\bibitem[Vasimuddin et~al\mbox{.}(2019)]%
        {bwa-mem2}
\bibfield{author}{\bibinfo{person}{Md. Vasimuddin}, \bibinfo{person}{Sanchit Misra}, \bibinfo{person}{Heng Li}, {and} \bibinfo{person}{Srinivas Aluru}.} \bibinfo{year}{2019}\natexlab{}.
\newblock \showarticletitle{Efficient Architecture-Aware Acceleration of BWA-MEM for Multicore Systems}. In \bibinfo{booktitle}{\emph{2019 IEEE International Parallel and Distributed Processing Symposium (IPDPS)}}. \bibinfo{pages}{314--324}.
\newblock
\urldef\tempurl%
\url{https://doi.org/10.1109/IPDPS.2019.00041}
\showDOI{\tempurl}


\bibitem[Virgin and Todd(2011)]%
        {virgin2011metagenomics}
\bibfield{author}{\bibinfo{person}{Herbert~W Virgin} {and} \bibinfo{person}{John~A Todd}.} \bibinfo{year}{2011}\natexlab{}.
\newblock \showarticletitle{Metagenomics and personalized medicine}.
\newblock \bibinfo{journal}{\emph{Cell}} \bibinfo{volume}{147}, \bibinfo{number}{1} (\bibinfo{year}{2011}), \bibinfo{pages}{44--56}.
\newblock


\bibitem[Wu et~al\mbox{.}(2024)]%
        {wu2024abakus}
\bibfield{author}{\bibinfo{person}{Lingxi Wu}, \bibinfo{person}{Minxuan Zhou}, \bibinfo{person}{Weihong Xu}, \bibinfo{person}{Ashish Venkat}, \bibinfo{person}{Tajana Rosing}, {and} \bibinfo{person}{Kevin Skadron}.} \bibinfo{year}{2024}\natexlab{}.
\newblock \showarticletitle{Abakus: Accelerating k-mer Counting with Storage Technology}.
\newblock \bibinfo{journal}{\emph{ACM Transactions on Architecture and Code Optimization}} \bibinfo{volume}{21}, \bibinfo{number}{1} (\bibinfo{year}{2024}), \bibinfo{pages}{1--26}.
\newblock


\bibitem[Xu et~al\mbox{.}(2024)]%
        {xu2024performance}
\bibfield{author}{\bibinfo{person}{Jiexiong Xu}, \bibinfo{person}{Yue Qiu}, \bibinfo{person}{Yiquan Chen}, \bibinfo{person}{Yijing Wang}, \bibinfo{person}{Wenhai Lin}, \bibinfo{person}{Yiquan Lin}, \bibinfo{person}{Shushu Zhao}, \bibinfo{person}{Yuqi Liu}, \bibinfo{person}{Ying Wang}, {and} \bibinfo{person}{Wenzhi Chen}.} \bibinfo{year}{2024}\natexlab{}.
\newblock \showarticletitle{Performance Characterization of SmartNIC NVMe-over-Fabrics Target Offloading}. In \bibinfo{booktitle}{\emph{Proceedings of the 17th ACM International Systems and Storage Conference}}. \bibinfo{pages}{14--24}.
\newblock


\bibitem[Yan et~al\mbox{.}(2019)]%
        {yan2019alleviating}
\bibfield{author}{\bibinfo{person}{Mingyu Yan}, \bibinfo{person}{Xing Hu}, \bibinfo{person}{Shuangchen Li}, \bibinfo{person}{Abanti Basak}, \bibinfo{person}{Han Li}, \bibinfo{person}{Xin Ma}, \bibinfo{person}{Itir Akgun}, \bibinfo{person}{Yujing Feng}, \bibinfo{person}{Peng Gu}, \bibinfo{person}{Lei Deng}, {et~al\mbox{.}}} \bibinfo{year}{2019}\natexlab{}.
\newblock \showarticletitle{Alleviating irregularity in graph analytics acceleration: A hardware/software co-design approach}. In \bibinfo{booktitle}{\emph{Proceedings of the 52nd Annual IEEE/ACM International Symposium on Microarchitecture}}. \bibinfo{pages}{615--628}.
\newblock


\bibitem[Zerbino and Birney(2008)]%
        {zerbino2008velvet}
\bibfield{author}{\bibinfo{person}{Daniel~R Zerbino} {and} \bibinfo{person}{Ewan Birney}.} \bibinfo{year}{2008}\natexlab{}.
\newblock \showarticletitle{Velvet: algorithms for de novo short read assembly using de Bruijn graphs}.
\newblock \bibinfo{journal}{\emph{Genome research}} \bibinfo{volume}{18}, \bibinfo{number}{5} (\bibinfo{year}{2008}), \bibinfo{pages}{821--829}.
\newblock


\bibitem[Zhang et~al\mbox{.}(2024)]%
        {zhang2024harp}
\bibfield{author}{\bibinfo{person}{Yichi Zhang}, \bibinfo{person}{Dibei Chen}, \bibinfo{person}{Gang Zeng}, \bibinfo{person}{Jianfeng Zhu}, \bibinfo{person}{Zhaoshi Li}, \bibinfo{person}{Longlong Chen}, \bibinfo{person}{Shaojun Wei}, {and} \bibinfo{person}{Leibo Liu}.} \bibinfo{year}{2024}\natexlab{}.
\newblock \showarticletitle{Harp: Leveraging Quasi-Sequential Characteristics to Accelerate Sequence-to-Graph Mapping of Long Reads}. In \bibinfo{booktitle}{\emph{Proceedings of the 29th ACM International Conference on Architectural Support for Programming Languages and Operating Systems, Volume 3}}. \bibinfo{pages}{512--527}.
\newblock


\bibitem[Zhou et~al\mbox{.}(2021)]%
        {zhou2021ultra}
\bibfield{author}{\bibinfo{person}{Minxuan Zhou}, \bibinfo{person}{Lingxi Wu}, \bibinfo{person}{Muzhou Li}, \bibinfo{person}{Niema Moshiri}, \bibinfo{person}{Kevin Skadron}, {and} \bibinfo{person}{Tajana Rosing}.} \bibinfo{year}{2021}\natexlab{}.
\newblock \showarticletitle{Ultra efficient acceleration for de novo genome assembly via near-memory computing}. In \bibinfo{booktitle}{\emph{2021 30th International Conference on Parallel Architectures and Compilation Techniques (PACT)}}. IEEE, \bibinfo{pages}{199--212}.
\newblock


\bibitem[Zhou et~al\mbox{.}(2023)]%
        {zhou2023dimm}
\bibfield{author}{\bibinfo{person}{Zhe Zhou}, \bibinfo{person}{Cong Li}, \bibinfo{person}{Fan Yang}, {and} \bibinfo{person}{Guangyu Sun}.} \bibinfo{year}{2023}\natexlab{}.
\newblock \showarticletitle{DIMM-Link: Enabling efficient inter-dimm communication for near-memory processing}. In \bibinfo{booktitle}{\emph{2023 IEEE International Symposium on High-Performance Computer Architecture (HPCA)}}. IEEE, \bibinfo{pages}{302--316}.
\newblock


\end{thebibliography}



\end{document}